\documentclass[twocolumn,english,aps,prl,superscriptaddress,address,amsmath,showpacs]{revtex4-1}
\usepackage[T1]{fontenc}
\usepackage[latin9]{inputenc}
\usepackage{color}
\usepackage{babel}
\usepackage{units}
\usepackage{amsmath}
\usepackage{amssymb}
\usepackage{graphicx}
\usepackage{esint}
\usepackage[unicode=true,pdfusetitle,
 bookmarks=true,bookmarksnumbered=false,bookmarksopen=false,
 breaklinks=false,pdfborder={0 0 0},backref=false,colorlinks=true]
 {hyperref}
\hypersetup{
 linkcolor=black}
\usepackage{breakurl}

\makeatletter

\usepackage[color=green!70,colorinlistoftodos]{todonotes}

\usepackage{natbib}
\usepackage{url}
\usepackage[amssymb]{SIunits}

\makeatother

\begin{document}

\title{Dynamics of a dipolar Bose-Einstein condensate in the vicinity of
a superconductor}

\author{Igor Sapina}

\affiliation{Universit\"at Bielefeld, Fakult\"at f\"ur Physik, Postfach 100131, D-33501
Bielefeld, Germany}

\affiliation{Institut f\"ur Theoretische Physik and Center for Collective Quantum
Phenomena, Universit\"at T\"ubingen, Auf der Morgenstelle 14, D-72076
T\"ubingen, Germany}

\author{Thomas Dahm}

\affiliation{Universit\"at Bielefeld, Fakult\"at f\"ur Physik, Postfach 100131, D-33501
Bielefeld, Germany}

\affiliation{Institut f\"ur Theoretische Physik and Center for Collective Quantum
Phenomena, Universit\"at T\"ubingen, Auf der Morgenstelle 14, D-72076
T\"ubingen, Germany}

\date{\today}
\begin{abstract}
We study the dynamics of a dipolar Bose-Einstein condensate, like
for example a $^{52}$Cr or $^{164}$Dy condensate, interacting with
a superconducting surface. The magnetic dipole moments of the atoms
in the Bose-Einstein condensate induce eddy currents in the superconductor.
The magnetic field generated by eddy currents modifies the trapping
potential such that the center-of-mass oscillation frequency is shifted.
We numerically solve the Gross-Pitaevskii equation for this system
and compare the results with analytical approximations. We present
an approximation that gives excellent agreement with the numerical
results. The eddy currents give rise to anharmonic terms, which leads
to the excitation of shape fluctuations of the condensate. We discuss
how the strength of the excitation of such modes can be increased
by exploiting resonances, and we examine the strength of the resonances
as a function of the center-of-mass oscillation amplitude of the condensate.
Finally, we study different
orientations of the magnetic dipoles and discuss favorable conditions
for the experimental observation of the eddy current effect. 
\end{abstract}


\maketitle

\section{Introduction}

Magnetic microtraps are a versatile tool to trap and manipulate
ultracold atomic gases and Bose-Einstein condensates (BEC).
Such traps provide a strong confinement and can be integrated
on a chip allowing the creation of specialized potentials
and control of ultracold atomic gases by electronic means
\cite{FortaghZimmermann}.
Also, the interaction of ultracold atomic gases with the
surface of the nearby solid can be studied. However, the
normal conductors that create the trapping potential at the
same time also create noise radiation from current fluctuations,
which limits the lifetime of the atomic cloud when it is brought
close to the conductor \cite{Jones,Rekdal}.
Recently, microtraps using superconductors have been realized
\cite{Nirrengarten,Mukai,CanoPRL,Siercke,Zhang,Mukai2014}.
In such microtraps the noise due to current fluctuations
is significantly suppressed in the relevant frequency range
due to the energy gap of the superconductor. This allows lifetimes several
orders of magnitudes longer than in conventional microtraps
\cite{Kasch,Skagerstam,Hohenester,Fermani}.
Such superconducting microtraps allow studying fundamental
interactions between BECs and superconductors and promise
coupling of these two macroscopic quantum phenomena
\cite{Fleischhauer,Verdu,Folman,Fischer,Bensky,Salem,CanoEPJD,Bernon,Cirac,Bothner}.

A disadvantage of superconducting microtraps is the
screening of the magnetic trapping fields due to
the Meissner effect, which has been shown to
lower the trap depth \cite{CanoPRL,CanoPRA}.
However, theoretical calculations have shown that
in spite of the Meissner effect distances below 1~$\mu$m 
can be achieved with superconducting
microtraps, if the edge enhancement of the currents in microstrips
of rectangular cross-section is exploited \cite{Markowsky,Dikovsky,Sokolovsky}.
A recent experiment has demonstrated a magnetic microtrap at a distance
of 14~$\mu$m from a superconductor \cite{Bernon}.

In the present work we study the interaction between a dipolar Bose-Einstein condensate
\cite{Lahaye2009} and a superconducting surface.  
Dipolar BECs consist of atoms which carry a large magnetic dipole moment. 
The first experimental realization of a dipolar BEC succeeded with $^{52}$Cr \cite{Chrom1}. 
Recently also the condensation of atoms with even larger dipole moments, like $^{168}$Er \cite{Erbium} or $^{164}$Dy \cite{Dysprosium} was reported.
Theses systems can be used to study a number of different properties \cite{Erbium2014,Dysprosium2014}. In the present work we consider center-of-mass oscillations of a dipolar BEC perpendicular to a superconducting surface. 
 The magnetic field, generated by the dipoles, induces eddy
currents in the superconducting surface. The eddy currents generate
a magnetic field, which in turn influences the BEC. This back action
on the BEC causes a shift of the center-of-mass oscillation frequency
relative to the case without a superconducting surface. In a previous
work \cite{Sapina2013} we have shown that this eddy current effect
generates a frequency shift which can be large enough to be detected
experimentally. The characteristic dependence of the frequency shift
as a function of the number of atoms in the BEC provides a fingerprint
which allows to identify this eddy current effect and separate it
from other surface effects like the Casimir-Polder force \cite{Harber,Antezza2004},
for example. We calculated the frequency shift using a relatively
simple column density model for the BEC, which allowed us to find
an analytical approximation. Furthermore the anharmonicity generated
by the BEC-surface potential leads to a coupling of the center-of-mass
motion to other collective modes of the BEC.

In our previous work we have made several simplifications: we have
neglected the influence of the dipole-dipole interaction on the dynamics
of the BEC, we used an effective anharmonic potential to emulate the
effect of the superconducting surface, and we considered only small
amplitude oscillations of the condensate. In the present work we will
present a more complete and accurate investigation of the effect.
We use a more realistic model by numerically solving the Gross-Pitaevskii
equation (GPE) for a dipolar BEC and including the full potential
generated by the surface. We will show that our previous results remain
qualitatively unchanged with some quantitative corrections. Furthermore,
we will consider larger amplitude oscillations. In addition, we study
different polarization directions of the condensate and find another
characteristic feature which can be used to experimentally identify
the eddy current effect. In addition to the resonant excitation of
the breather mode, on which we reported in \cite{Sapina2013}, we
find the resonant excitation of a different collective mode, which
did not appear in our previously used effective model.

\begin{figure}
\includegraphics[width=1\columnwidth]{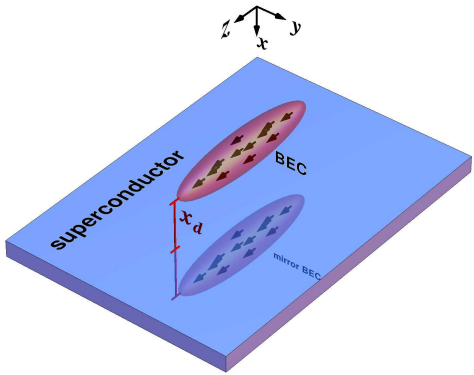}

\caption{\label{fig:BEC_and_mirror} (Color online) Depicted is a schematic sketch of the system
under investigation. A Bose-Einstein condensate is placed in a distance
$x_{d}$ above a superconducting surface. The BEC consists of atoms
which carry a magnetic dipole moment. The dipoles are all aligned
in the same direction by a magnetic field. The magnetic field generated
by the dipoles must satisfy the boundary condition $\mathbf{B}\cdot\hat{\mathbf{n}}=0$
at the surface of the superconductor, where $\hat{\mathbf{n}}$ is
the normal vector of the superconducting plane. By introducing a magnetic
mirror BEC, in a distance $x_{d}$ below the superconducting surface,
this boundary condition can be satisfied. With that the interaction
between BEC and superconductor can be modeled as interaction between
BEC and mirror BEC.}

\end{figure}

\section{Numerical solution of the Gross-Pitaevskii Equation for a dipolar BEC close to a superconducting surface}

\subsection{Modeling the system and the numerical approach}

\subsubsection{The Gross-Pitaevskii equation of a dipolar BEC close to a superconducting
surface}

Here we present the investigated system and explain the model that
we use for our calculations. We consider a dilute gas of Bose atoms
trapped in a harmonic potential. The potential can be generated by
optical or magnetic means. For the setup under consideration, a superconducting
microtrap might be the most convenient choice. The temperature of
the gas shall be cooled far below the transition temperature where
Bose-Einstein condensation occurs and we assume that the temperature
of the gas is $T=0$. This means that all the atoms in the trap will
be in the condensate. Every atom carries a magnetic dipole moment
$\mathbf{m}$. The dipoles are all aligned in the direction of an
external magnetic field. The many body Hamiltonian for this Bose gas
reads
\begin{equation}
\hat{H}=\sum_{i=1}^{N}\left(\frac{\mathbf{p}_{i}^{2}}{2M}+V_{T}\left(\mathbf{r}_{i}\right)\right)+\frac{1}{2}\sum_{i=1}^{N}\sum_{j\neq i}^{N}U\left(\mathbf{r}_{i},\mathbf{r}_{j}\right),\label{eq:many_body_hamiltonian_dipolar_BEC}
\end{equation}
where $U\left(\mathbf{r}_{i},\mathbf{r}_{j}\right)$ is the interaction
potential between two atoms. The atoms can interact via short ranged
$s$-wave interaction
\begin{equation}
U_{s}\left(\mathbf{r},\mathbf{r}^{\prime}\right)=g_{s}\delta^{(3)}\left(\mathbf{r}-\mathbf{r}^{\prime}\right)\label{eq:s-wave-scattering_potential}
\end{equation}
and via long ranged dipole-dipole interaction
\begin{equation}
U_{\mathrm{md}}\left(\mathbf{r},\mathbf{r}^{\prime}\right)=-\,\frac{\mu_{0}}{4\pi}\left(\frac{3\left(\mathbf{m}\cdot\hat{\mathbf{n}}\right)\left(\mathbf{m}^{\prime}\cdot\hat{\mathbf{n}}\right)-\mathbf{m}\cdot\mathbf{m}^{\prime}}{\left|\mathbf{r}-\mathbf{r}^{\prime}\right|^{3}}\right),\label{eq:dipole-dipole-interaction_potential}
\end{equation}
where $\hat{\mathbf{n}}=\frac{\mathbf{r}-\mathbf{r}^{\prime}}{\left|\mathbf{r}-\mathbf{r}^{\prime}\right|}$
is the normalized distance vector, $\mathbf{m}$ and $\mathbf{m}^{\prime}$
are the magnetic dipole moments of the two interacting dipoles. The
external trapping potential is given by 
\begin{equation}
V_{T}\left(\mathbf{r}\right)=\frac{M}{2}\left(\omega_{x}^{2}x^{2}+\omega_{y}^{2}y^{2}+\omega_{z}^{2}z^{2}\right),\label{eq:harmonic_potential}
\end{equation}
where $M$ is the atom mass and $\omega_{x}$, $\omega_{y}$ and $\omega_{z}$
are the trapping frequencies. With the Hartree ansatz for the many
body wave function
\begin{equation}
\Psi_{H}\left(\mathbf{r}_{1},\mathbf{r}_{2},\ldots\mathbf{r}_{N}\right)=\psi\left(\mathbf{r}_{1}\right)\psi\left(\mathbf{r}_{2}\right)\ldots\psi\left(\mathbf{r}_{N}\right)\label{eq:hartree_ansatz}
\end{equation}
the energy functional $E=\left\langle \Psi_{H}\right|\hat{H}\left|\Psi_{H}\right\rangle $
can be minimized with respect to $\psi$ under the constraint $\left\langle \Psi_{H}\right.\left|\Psi_{H}\right\rangle =1$,
which then yields the Gross-Pitaevskii equation \cite{Pethick&Smith,Pitaevskii&Stingari}
\begin{eqnarray}
\mu\psi\left(\mathbf{r}\right) & = & \Bigg(-\frac{\hbar^{2}}{2M}\boldsymbol{\nabla}^{2}+V\left(\mathbf{r}\right)\label{eq:stationary_GPE}\\
\nonumber \\
 &  & +\left(N-1\right)\intop_{\mathbb{R}^{3}}\mathrm{d}\mathbf{r}^{\prime}\, U\left(\mathbf{r},\mathbf{r}^{\prime}\right)\left|\psi\left(\mathbf{r}^{\prime}\right)\right|^{2}\Bigg)\psi\left(\mathbf{r}\right).\nonumber 
\end{eqnarray}
 So far, this is the standard way to describe a dipolar BEC. Next
we need to include the interaction with the superconducting surface.
As described in \cite{Sapina2013} the magnetic interaction with an
infinitely extended superconducting surface can be modeled by the
interaction of the BEC with its magnetic mirror, as is depicted in
Fig. \ref{fig:BEC_and_mirror}. Every atom in the BEC interacts with
every atom in the mirror BEC via magnetic dipole-dipole interaction.
We have to add
\[
\hat{H}_{\mathrm{mir}}=\sum_{i=1}^{N}\sum_{k=1}^{N}U_{\mathrm{md}}\left(\mathbf{r}_{i},\mathbf{r}_{k}^{\prime}\right)
\]
to the many body Hamiltonian (\ref{eq:many_body_hamiltonian_dipolar_BEC}).
The index $i$ denotes the particles in the BEC and the index $k$
the mirror particles. Instead of the usual Hartree ansatz (\ref{eq:hartree_ansatz}),
we now make the ansatz
\begin{equation}
\Phi_{H}\left(\mathbf{r}_{1},\ldots\mathbf{r}_{N};\mathbf{r}_{1}^{\prime},\ldots\mathbf{r}_{N}^{\prime}\right)=\prod_{i=1}^{N}\psi\left(\mathbf{r}_{i}\right)\prod_{k=1}^{N}\chi\left(\mathbf{r}_{k}^{\prime}\right).\label{eq:hartree_ansatz_mixture}
\end{equation}
It is the Hartree ansatz for a BEC consisting of a mixture of two
different kinds of bosons. In our case, $\psi\left(\mathbf{r}_{i}\right)$
is the single body wave function of an atom in the BEC, while $\chi\left(\mathbf{r}_{k}^{\prime}\right)$
describes an atom in the mirror BEC. Theses two kinds of atoms can
interact via long ranged dipole-dipole interaction. Also, $\chi$
is not an independent function. It is the shifted mirror function
of $\psi$. Now the task is to minimize 
\[
E-\mu N=\left\langle \Phi_{H}\right|\hat{H}_{0}\left|\Phi_{H}\right\rangle +\left\langle \Phi_{H}\right|\hat{H}_{\mathrm{mir}}\left|\Phi_{H}\right\rangle -\mu\left\langle \Phi_{H}\right.\left|\Phi_{H}\right\rangle 
\]
with respect to the single body wave function $\psi$. This is presented
in Appendix \ref{sec:appendix:mirror_term_GPE}. The resulting GPE
reads
\begin{eqnarray}
\mu\cdot\psi\left(\mathbf{r}\right) & = & \Bigg(-\frac{\hbar^{2}}{2M}\boldsymbol{\nabla}^{2}+V_{T}\left(\mathbf{r}\right)+N\int\mathrm{d\mathbf{r}}^{\prime}\, U\left(\mathbf{r},\mathbf{r}^{\prime}\right)\left|\psi\left(\mathbf{r}^{\prime}\right)\right|^{2}\nonumber \\
\nonumber \\
 &  & +2N\intop\mathrm{d}\mathbf{r}^{\prime}\, U_{\mathrm{md}}\left(\mathbf{r},\mathbf{r}^{\prime}\right)\left|\chi\left(\mathbf{r}^{\prime}\right)\right|^{2}\Bigg)\cdot\psi\left(\mathbf{r}\right)\label{eq:GPE_with_mirror_term}
\end{eqnarray}

\subsubsection{Calculation of the mirror term}

The mirror term in (\ref{eq:GPE_with_mirror_term}) introduces an
additional complication into solving this non-linear Schroedinger
equation. Next we want to present an efficient way to calculate this
term. First of all we assume that the potential generated by the mirror
BEC is small compared to the trapping potential $V_{T}\left(\mathbf{r}\right)$,
as well as to the interaction strength between the atoms. If that
is the case, then small deviations from the exact shape of $\left|\chi\left(\mathbf{r}^{\prime}\right)\right|^{2}$
will not be significant. As is well known, if the interaction between
the atoms becomes large enough, the kinetic term in the GPE can be
neglected. This leads to the so-called Thomas-Fermi approximation
\cite{Pethick&Smith}. Within this approximation the density distribution
of a BEC in a harmonic potential takes the guise of an ellipsoid
\begin{equation}
n_{\mathrm{TF}}\left(\mathbf{r}\right)=n_{0}\left(1-\frac{x^{2}}{\lambda_{x}^{2}}-\frac{y^{2}}{\lambda_{y}^{2}}-\frac{z^{2}}{\lambda_{z}^{2}}\right),\label{eq:TF-density}
\end{equation}
where $\lambda_{x}$, $\lambda_{y}$ and $\lambda_{z}$ are the semi-axes
of the ellipsoid and $n_{0}$ is the central density. In the case
that there is no dipole-dipole interaction present between the atoms,
the semi-axes are given by 
\begin{equation}
\lambda_{a}^{(0)}=\sqrt{\frac{2\mu}{m\omega_{a}^{2}}}.
\label{eq:TF-semi-axes-ed=0}
\end{equation}
In the presence of dipole-dipole interaction the semi-axes are modified
\cite{Eberlein}. They need to be determined numerically from a set
of coupled self-consistency equations \cite{Sapina}. The central
density $n_{0}$ can be determined from the requirement $N=\intop\mathrm{d}\mathbf{r}^{\prime}\, n\left(\mathbf{r}^{\prime}\right)$,
and is given by $n_{0}=\frac{15}{8\pi}\frac{N}{\lambda_{x}\lambda_{y}\lambda_{z}}$.

\begin{figure}
\includegraphics[width=1\columnwidth]{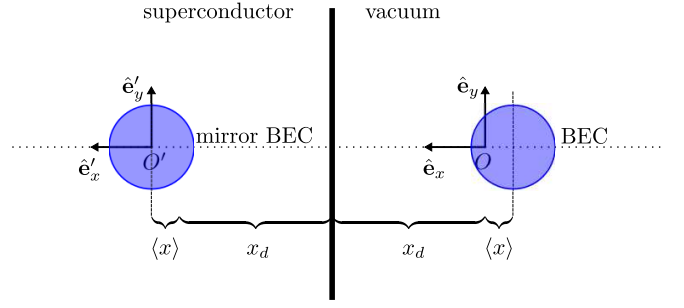}

\caption{(Color online) Depicted are the relative positions of the BEC and its mirror. The
coordinate system $K^{\prime}$ is co-moving with the mirror BEC and
has its origin $O^{\prime}$ at the center of the mirror BEC. The
coordinate system $K$ is stationary and its origin $O$ coincides
with the minimum of the harmonic trap. As becomes clear from this
graphic, the transformation between $K^{\prime}$ and $K$ is given
by $x=x^{\prime}-2x_{d}-\left\langle x\right\rangle $.}

\label{fig:coordinate_systems}
\end{figure}

In the following we will use the Thomas-Fermi approximation to model
the mirror BEC. But first let us rewrite $U_{\mathrm{md}}\left(\mathbf{r},\mathbf{r}^{\prime}\right)$.
In the case that the external polarizing magnetic field is oriented
in the $z$-direction we have $\mathbf{m}=\mathbf{m}^{\prime}=m\cdot\hat{\mathbf{e}}_{z}$
and with that
\begin{equation}
U_{\mathrm{md}}\left(\mathbf{r},\mathbf{r}^{\prime}\right)=-\frac{g_{D}}{4\pi}\left(\frac{3\left(z-z^{\prime}\right)^{2}}{\left|\mathbf{r}-\mathbf{r}^{\prime}\right|^{5}}-\frac{1}{\left|\mathbf{r}-\mathbf{r}^{\prime}\right|^{3}}\right).\label{eq:magnetic-dipole-dipole-interaction}
\end{equation}
A more convenient way to write this potential is
\begin{equation}
U_{\mathrm{md}}\left(\mathbf{r},\mathbf{r}^{\prime}\right)=-\frac{g_{D}}{4\pi}\left(\frac{\partial^{2}}{\partial z^{2}}\frac{1}{\left|\mathbf{r}-\mathbf{r}^{\prime}\right|}+\frac{4\pi}{3}\delta\left(\mathbf{r}-\mathbf{r}^{\prime}\right)\right),\label{eq:magnetic-dipole-dipole-interaction_dipoles-in-z-direction}
\end{equation}
the first term represents the long ranged part of the interaction
and the second term the short ranged part. Since we calculate the
interaction between the BEC and its mirror, the delta distribution
can never contribute to $U_{\mathrm{md}}\left(\mathbf{r},\mathbf{r}^{\prime}\right)$,
so for $N\left|\chi\left(\mathbf{r}^{\prime}\right)\right|^{2}=n_{\mathrm{TF}}\left(\mathbf{r}^{\prime}\right)$
we get 
\begin{eqnarray*}
V_{\mathrm{mir}}\left(\mathbf{r}\right) & = & N\intop_{\mathbb{D}_{\mathrm{TF}}}\mathrm{d}\mathbf{r}^{\prime}\, U_{\mathrm{md}}\left(\mathbf{r},\mathbf{r}^{\prime}\right)\left|\chi\left(\mathbf{r}^{\prime}\right)\right|^{2}\\
\\
 & = & -g_{D}\frac{\partial^{2}}{\partial z^{2}}\frac{1}{4\pi}\intop_{\mathbb{D}_{\mathrm{TF}}}\mathrm{d}\mathbf{r}^{\prime}\,\frac{n_{\mathrm{TF}}\left(\mathbf{r}^{\prime}\right)}{\left|\mathbf{r}-\mathbf{r}^{\prime}\right|}\\
\\
 & = & -g_{D}n_{0}\frac{\partial^{2}}{\partial z^{2}}\phi\left(\mathbf{r}\right),
\end{eqnarray*}
with
\[
\mathbb{D}_{\mathrm{TF}}=\left\{ \mathbf{r}\in\mathbb{R}^{3}\left|\frac{x^{2}}{\lambda_{x}^{2}}+\frac{y^{2}}{\lambda_{y}^{2}}+\frac{z^{2}}{\lambda_{z}^{2}}\leq1\right.\right\} ,
\]
and 
\begin{equation}
\phi\left(\mathbf{r}\right)=\frac{1}{4\pi}\intop_{\mathbb{D}_{\mathrm{TF}}}\mathrm{d}\mathbf{r}^{\prime}\,\frac{1}{\left|\mathbf{r}-\mathbf{r}^{\prime}\right|}\left(1-\frac{x^{\prime2}}{\lambda_{x}^{2}}-\frac{y^{\prime2}}{\lambda_{y}^{2}}-\frac{z^{\prime2}}{\lambda_{z}^{2}}\right).\label{eq:chandrasekhar_integral}
\end{equation}
With that we have reformulated the task into determining the potential
function $\phi\left(\mathbf{r}\right)$. Formally, this is the same
task as to determine the gravitational potential of an ellipsoidal
mass distribution. Chandrasekhar provides a detailed discussion of
this type of elliptic integrals in the context of rotating gas clouds
\cite{Chandrasekhar}. He presents an exact one dimensional representation
for $\phi\left(\mathbf{r}\right)$, for the case that $\mathbf{r}\in\mathbb{D}_{\mathrm{TF}}$
as well as for $\mathbf{r}\notin\mathbb{D}_{\mathrm{TF}}$. The case
$\mathbf{r}\in\mathbb{D}_{\mathrm{TF}}$ is useful if one is interested
in calculating the potential between the atoms in the BEC. For example
to calculate the semi-axes of a dipolar BEC \cite{Sapina} or its
collective modes \cite{Sapina,vanBijnen}. Since we want to calculate
the potential of the mirror cloud at the position of the actual BEC
we need the case $\mathbf{r}\notin\mathbb{D}_{\mathrm{TF}}$. In this
case the one-dimensional representation of $\phi\left(\mathbf{r}\right)$
reads
\begin{equation}
\phi\left(\mathbf{r}\right)=\frac{\lambda_{x}\lambda_{y}\lambda_{z}}{8}\intop_{W\left(\mathbf{r}\right)}^{\infty}\mathrm{d}u\frac{\left(1-\frac{x^{2}}{\lambda_{x}^{2}+u}-\frac{y^{2}}{\lambda_{y}^{2}+u}-\frac{z^{2}}{\lambda_{z}^{2}+u}\right)^{2}}{\sqrt{\left(\lambda_{x}^{2}+u\right)\left(\lambda_{y}^{2}+u\right)\left(\lambda_{z}^{2}+u\right)}}.\label{eq:chandrasekhar_integral_1D}
\end{equation}
The function $W\left(\mathbf{r}\right)$ is the ellipsoidal coordinate
of the point $\mathbf{r}$ and is defined by 
\begin{equation}
\frac{x^{2}}{\lambda_{x}^{2}+W\left(\mathbf{r}\right)}+\frac{y^{2}}{\lambda_{y}^{2}+W\left(\mathbf{r}\right)}+\frac{z^{2}}{\lambda_{z}^{2}+W\left(\mathbf{r}\right)}=1.\label{eq:ellipsoidal_coordinate_W}
\end{equation}
In the case $\mathbf{r}\in\mathbb{D}_{\mathrm{TF}}$ the lower integration
limit of this integral would be $0$. In order to calculate the mirror
potential $V_{\mathrm{mir}}\left(\mathbf{r}\right)$ we need the second
derivative of $\phi\left(\mathbf{r}\right)$ with respect to $z$.
The detailed calculation can be found in appendix \ref{sec:appendix:calculation_mirror_term},
here we only present the result
\begin{eqnarray*}
\frac{\partial^{2}\phi\left(\mathbf{r}\right)}{\partial z^{2}} & = & -\frac{\lambda_{x}\lambda_{y}\lambda_{z}}{2}\\
\\
 &  & \times\intop_{W\left(\mathbf{r}\right)}^{\infty}\mathrm{d}u\,\frac{\left(1-\frac{x^{2}}{\lambda_{x}^{2}+u}-\frac{y^{2}}{\lambda_{y}^{2}+u}-3\frac{z^{2}}{\lambda_{z}^{2}+u}\right)}{\left(\lambda_{z}^{2}+u\right)\sqrt{\left(\lambda_{x}^{2}+u\right)\left(\lambda_{y}^{2}+u\right)\left(\lambda_{z}^{2}+u\right)}}.
\end{eqnarray*}
Following Ref. \cite{Sapina} we now introduce the index integrals
\begin{equation}
J_{a}\equiv J_{a}\left(\mathbf{r}\right)=\intop_{W\left(\mathbf{r}\right)}^{\infty}\frac{\mathrm{d}u}{\sqrt{\beta\left(u\right)}}\frac{1}{\left(\lambda_{a}^{2}+u\right)}\label{eq:GPE:index-integral-Ja}
\end{equation}
and
\begin{equation}
J_{ab}\equiv J_{ab}\left(\mathbf{r}\right)=\intop_{W\left(\mathbf{r}\right)}^{\infty}\frac{\mathrm{d}u}{\sqrt{\beta\left(u\right)}}\frac{1}{\left(\lambda_{a}^{2}+u\right)}\frac{1}{\left(\lambda_{b}^{2}+u\right)},\label{eq:GPE:index-integral-Jab}
\end{equation}
with $\beta\left(u\right)=\left(\lambda_{x}^{2}+u\right)\left(\lambda_{y}^{2}+u\right)\left(\lambda_{z}^{2}+u\right)$.
Using the index integrals we can rewrite $\frac{\partial^{2}}{\partial z^{2}}\phi\left(\mathbf{r}\right)$,
which then reads
\begin{equation}
\begin{split}\frac{\partial^{2}\phi\left(\mathbf{r}\right)}{\partial z^{2}}=\lambda_{x}\lambda_{y}\lambda_{z}\Bigg( & -\frac{1}{2}J_{z}+\frac{1}{2}J_{xz}\cdot x^{2}\\
 & +\frac{1}{2}J_{yz}\cdot y^{2}+\frac{3}{2}J_{zz}\cdot z^{2}\Bigg).
\end{split}
\label{eq:ddz_phi_with_index_integrals}
\end{equation}
In the numerical calculations the coordinate system $K$ is chosen
such that the origin coincides with the minimum of the harmonic potential
$V_{T}\left(\mathbf{r}\right)$. Expression (\ref{eq:ddz_phi_with_index_integrals})
only holds in $K^{\prime}$, which is the frame of reference where
the origin coincides with the center of the mirror BEC. The transformation
between the two Systems $K$ and $K^{\prime}$ is given by $x=x^{\prime}-2x_{d}-\left\langle x\right\rangle $,
as is depicted in Fig. \ref{fig:coordinate_systems}. $\left\langle x\right\rangle $
is the $x$-coordinate of the BEC center-of-mass calculated in $K$.
Since the BEC oscillates in the $x$-direction, $\left\langle x\right\rangle $
is a function of time. While $K$ is a stationary frame of reference,
$K^{\prime}$ is co-moving with the mirror BEC opposite to the motion
of the BEC. Expressed in $K$ the mirror potential takes the guise
 
\begin{equation}
\tilde{V}_{\mathrm{mir}}\left(\mathbf{r};\left\langle x\right\rangle \right)=-g_{D}n_{0}\left.\frac{\partial^{2}\phi\left(\mathbf{r}^{\prime}\right)}{\partial z^{\prime2}}\right|_{\mathbf{r}^{\prime}=\mathbf{r}+\left(2x_{d}+\left\langle x\right\rangle \right)\hat{\mathbf{e}}_{x}}.\label{eq:V_mir}
\end{equation}
Note, that the index integrals, which appear in (\ref{eq:ddz_phi_with_index_integrals}),
also depend on the position $\mathbf{r}$ via the lower integration
limit $W\left(\mathbf{r}\right)$. This means that $V_{\mathrm{mir}}$
is not simply a quadratic form. Since the BEC is in motion, $\left\langle x\right\rangle $
is a time dependent quantity, which makes $\tilde{V}_{\mathrm{mir}}$
a time dependent potential.

We have now reduced the three dimensional integral in the GPE (\ref{eq:GPE_with_mirror_term})
to four one dimensional index integrals. For a numerical calculation
this is already a huge advantage. As it turns out, it is not even
necessary to calculate all four integrals, since there exist algebraic
relations between the integrals $J_{a}$ and $J_{ab}$ which can be
exploited. This is shown in Appendix \ref{sec:appendix:the_index_integrals}.

\subsubsection{Time evolution of the BEC}

In order to calculate the time evolution of the BEC we need to solve
the time dependent GPE. After all that has been said, the time dependent
GPE reads
\begin{widetext}
\begin{equation}
\begin{split}i\hbar\frac{\partial}{\partial t}\psi\left(\mathbf{r},t\right)=\Bigg( & -\frac{\hbar^{2}}{2M}\boldsymbol{\nabla}^{2}+\frac{M}{2}\left(\omega_{x}^{2}x^{2}+\omega_{y}^{2}y^{2}+\omega_{z}^{2}z^{2}\right)+Ng_{s}\Bigg[\left(1-\varepsilon_{D}\right)\left|\psi\left(\mathbf{r},t\right)\right|^{2}\\
\\
 & -3\varepsilon_{D}\frac{1}{4\pi}\frac{\partial^{2}}{\partial z^{2}}\int\mathrm{d\mathbf{r}}^{\prime}\frac{\left|\psi\left(\mathbf{r}^{\prime},t\right)\right|^{2}}{\left|\mathbf{r}-\mathbf{r}^{\prime}\right|}-\varepsilon_{D}^{(m)}\frac{45}{4\pi\lambda_{x}\lambda_{y}\lambda_{z}}\left.\frac{\partial^{2}\phi\left(\mathbf{r}^{\prime}\right)}{\partial z^{\prime2}}\right|_{\mathbf{r}^{\prime}=\mathbf{r}+\left(2x_{d}+\left\langle x\right\rangle \right)\hat{\mathbf{e}}_{x}}\Bigg]\Bigg)\cdot\psi\left(\mathbf{r},t\right).
\end{split}
\label{eq:time_dep_GPE}
\end{equation}
\end{widetext}
Here we have introduced the dipole-dipole interaction
parameter
\begin{equation}
\varepsilon_{D}=\frac{g_{D}}{3g_{s}}.\label{eq:epsilon_d}
\end{equation}
It is a dimensionless parameter which measures the strength of the
dipole-dipole interaction relative to the strength of the contact
interaction. In a harmonic potential the stability of the ground state
is only guaranteed if $-1/2<\varepsilon_{D}<1$. In the following
we will only consider values of $\varepsilon_{D}$ that reside in
the positive part of this interval. As can be seen from the GPE, the
parameter $\varepsilon_{D}$ reduces the strength of the contact interaction
and introduces a long ranged interaction between the atoms. In order
to distinguish the interaction between the atoms in the BEC and the
interaction with the superconducting surface, we introduce the parameter
$\varepsilon_{D}^{(m)}$. It is defined in the same way as $\varepsilon_{D}$.
The only difference is that it describes the interaction with the
mirror BEC. While in a real setup those two parameters will always
have the same value, we will sometimes choose $\varepsilon_{D}$ to
be zero, while $\varepsilon_{D}^{(m)}$ is non-zero.

Before the time evolution can be calculated, the ground state needs
to be determined. To do so, we need to solve the stationary GPE. For
this we use a backward Euler method \cite{Bao2010,Bao2004,Bao2006,Bao2013},
where we calculate the kinetic term using a Fourier transformation.
The long ranged part of the dipole-dipole interaction is also calculated
with a Fourier transformation. As an external parameter for the numerical
calculation we can set the distance between the superconductor and
the minimum of the harmonic trap $V_{T}\left(\mathbf{r}\right)$.
The actual distance between the BEC and the superconductor slightly
differs from this value. The reason for that is the mirror interaction
potential, it causes a shift of the minimum of the overall potential.
With that, the equilibrium position of the BEC is also shifted. The
BEC will not oscillate around the harmonic trap minimum, but around
this new potential minimum. However, this shift is so tiny that it
can not be detected in an experiment. For this reason we will not
discuss it here any further. Once the ground state is determined,
we shift the harmonic trap minimum by $x_{s}$ in the $x-$direction
to create an initial state for the oscillation of the condensate.
After the shift, the distance from the harmonic trap minimum to the
surface is $x_{d}$. We have now created an excited state, which performs
a center-of-mass oscillation around the potential minimum with amplitude
$x_{s}$. We compute the time evolution with a time-splitting spectral
method \cite{Bao2010,Bao2006,Bao2013,TSSP}. Again, the gradient term
and the long ranged dipole-dipole interaction potential are taken
care of by Fourier transformations. For the spatial discretization
of the BEC wave function we use a $64\times64\times64$ lattice with
periodic boundary conditions. To reduce the computing time, we parallelized
parts of the code necessary to calculate a single time step. Those
parallel parts of the code were computed on the GPU. We implemented
this using CUDA.

In each time step the potential generated by the mirror BEC needs
to be determined. The mirror potential is a function of the $x$-coordinate
of the center-of-mass position
\[
x_{n}\equiv\left\langle x\left(t_{n}\right)\right\rangle =\intop\mathrm{d}\mathbf{r}\, x\cdot\left|\psi\left(\mathbf{r},t_{n}\right)\right|^{2}.
\]
During a time step $\Delta t$ the position of the center-of-mass
shifts from a position $x_{n}$ to a position $x_{n+1}$. If we use
the position $x_{n}$ to calculate the mirror potential, we introduce
a systematic error into our calculation. To avoid this, we need a
method to calculate an effective center-of-mass position for the whole
time step, like for example $x_{\mathrm{eff}}=\left(x_{n}+x_{n+1}\right)/2$.
To calculate $x_{\mathrm{eff}}$ we would need the wave function at
the end of the time step, which would require a self consistent calculation
of every time step. To avoid this, we make use of the time splitting
scheme for the time discretization. The Hamiltonian we use in (\ref{eq:stationary_GPE})
can be separated into two parts $\hat{H}=\hat{K}+\hat{V}$, with the
kinetic operator $\hat{K}=-\frac{\hbar^{2}}{2M}\boldsymbol{\nabla}^{2}$
and $\hat{V}$ everything else. We then decompose a single time step
from $t_{n}$ to $t_{n+1}$ via the Strang splitting method \cite{Strang},
where the time evolution operator is split in three parts. The wave
function $\psi_{n+1}\equiv\psi\left(\mathbf{r},t_{n+1}\right)$ can
be constructed from the wave function $\psi_{n}\equiv\psi\left(\mathbf{r},t_{n}\right)$
using the following scheme: 
\begin{eqnarray*}
\psi^{(1)} & = & \exp\left(i\frac{\hat{K}}{\hbar}\frac{\Delta t}{2}\right)\psi_{n},\\
\\
\psi^{(2)} & = & \exp\left(i\frac{\hat{V}}{\hbar}\Delta t\right)\psi^{(1)},\\
\\
\psi_{n+1} & = & \exp\left(i\frac{\hat{K}}{\hbar}\frac{\Delta t}{2}\right)\psi^{(2)}.
\end{eqnarray*}
As can be shown, the application of $\exp\left(i\frac{\hat{V}}{\hbar}\Delta t\right)$
does not change $\left|\psi\right|^{2}$. So the position of the center-of-mass
only changes after $\exp\left(i\frac{\hat{K}}{\hbar}\frac{\Delta t}{2}\right)$
has been applied. But since the mirror potential does not contribute
to $\hat{K}$, its shape and strength is not relevant for the first
part of the time step. After $\exp\left(i\frac{\hat{K}}{\hbar}\frac{\Delta t}{2}\right)$
has been applied, the position of the center-of-mass has shifted to
a value $\tilde{x}_{n}$. Before we now apply the operator $\exp\left(i\frac{\hat{V}}{\hbar}\Delta t\right)$,
we calculate the mirror potential using $\tilde{x}_{n}$ as the effective
center-of-mass position for the whole time step. We conclude the time
step by applying the operator $\exp\left(i\frac{\hat{K}}{\hbar}\frac{\Delta t}{2}\right)$
one more time, which shifts the center-of-mass to its final value
$x_{n+1}$.

Besides the $x$-coordinate of the center-of-mass we also keep track
of the widths
\begin{equation}
\sigma_{a}\left(t_{n}\right)=\sqrt{\left\langle \left(a-a_{n}\right)^{2}\right\rangle },\, a\in\left\{ x,y,z\right\} \label{eq:BEC-width}
\end{equation}
of the BEC. For a Thomas-Fermi ellipsoid, $\sigma_{a}$ is connected
to the semi-axes via $\sigma_{a}=\lambda_{a}/\sqrt{7}$. The analysis
of the respective time curves yields information about the excited
modes.

\subsection{\label{sub:Numerical_results_for_frequency_shift}Numerical results
for the center-of-mass frequency shift}

First we want to study the center-of-mass motion of the BEC. In ref.
\cite{Sapina2013}, where we followed the approach of Antezza et al.
\cite{Antezza2004}, we have used a simple column density model for
the BEC to calculate the frequency shift. This model had the advantage
that we were able to get some analytical results for the shift. Now
we want to compare the approximate results from the column density
model with the results of the numerical simulations. First of all
we expect to get good agreement for the case that the oscillation
amplitude $x_{s}$ is small compared to the BEC semi-axis $\lambda_{x}$.
But even in the case where $x_{s}\ll\lambda_{x}$, we have to expect
deviations due to the finite extension of the BEC in the $x$- and
$y$-direction, as the column density model is infinitively thin in
these directions.

\subsubsection{Small amplitude oscillations}

In order to improve agreement with the numerical results we can replace
the one dimensional column density distribution by a three dimensional
Thomas-Fermi density distribution. In the case of small amplitude
oscillations, the frequency shift is found to be
\begin{equation}
\gamma=\frac{\omega_{x}^{\prime}-\omega_{x}}{\omega_{x}}=\frac{1}{2\omega_{x}^{2}M}\frac{1}{N}\int\mathrm{d}\mathbf{r}\, n_{\mathrm{TF}}\left(\mathbf{r}\right)g\left(\mathbf{r};x_{d}\right),\label{eq:gamma_3D_Thomas-Fermi}
\end{equation}
where $\omega_{x}^{\prime}$ is the center-of-mass oscillation frequency
and $\omega_{x}$ is the harmonic trap frequency. The function $g\left(\mathbf{r};x_{d}\right)$
describes the curvature of the potential $V_{\mathrm{mir}}$, which
is generated by the mirror BEC
\begin{equation}
g\left(\mathbf{r};x_{d}\right)=4\left.\frac{\partial^{2}}{\partial x^{\prime2}}V_{\mathrm{mir}}\left(\mathbf{r}^{\prime}\right)\right|_{\mathbf{r}^{\prime}=\mathbf{r}+2x_{d}\hat{\mathbf{e}}_{x}}.\label{eq:curvature_due_to_mirror}
\end{equation}
A more detailed derivation of this result is found in Appendix \ref{sec:frequency_shift}.
The factor of $4$ in the function $g$ is due to the fact that the
mirror BEC moves opposite to the BEC. To account for this we need
to take the derivative with respect to $x/2$ rather than to $x$.
This leads to a factor of $4$ in the second derivative. In order
to calculate the function $g$, we again make use of index integrals.
The result is a rather long expression, so we will not give it here.
The integral which occurs in (\ref{eq:gamma_3D_Thomas-Fermi}) cannot
be further simplified by the use of index integrals, therefore we
calculate this three dimensional integral numerically. Here we see
the advantage of the column density model, instead of a three dimensional
integral, we have a one dimensional integral. The one dimensional
integral can be solved analytically, or if one prefers the one dimensional
integral can also be solved numerically, which involves just little
computational effort.

\begin{figure}
\includegraphics[width=1\columnwidth]{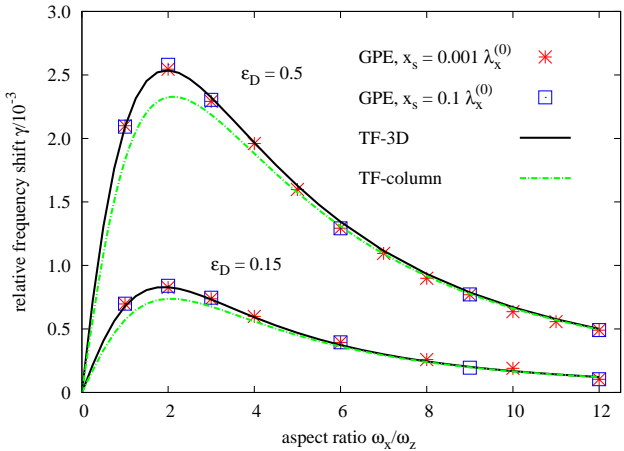}

\caption{\label{fig:frequency_shift_for_small_amplitudes}(Color online) Frequency shift for
small amplitude oscillations: We calculate the frequency shift for
two different dipole-dipole interaction strengths, $\varepsilon_{D}=\varepsilon_{D}^{(m)}=0.15$
and $\varepsilon_{D}=\varepsilon_{D}^{(m)}=0.5$. The data points
show the results from the numerical solution of the time dependent
GPE (\ref{eq:time_dep_GPE}). The red stars show the results for an
oscillation amplitude of $x_{s}=0.001\,\lambda_{x}^{(0)}$ and the
blue squares for $x_{s}=0.1\,\lambda_{x}^{(0)}$. The solid black
lines show the frequency shift based on the Thomas-Fermi approximation
for a three dimensional BEC for small amplitude oscillations (\ref{eq:gamma_3D_Thomas-Fermi}).
The dot-dashed green lines show the results based on the column density
model \cite{Sapina2013}. \emph{Other parameters}: $\kappa=\omega_{y}/\omega_{x}=1$,
distance to surface $x_{d}=2\lambda_{x}^{(0)}$, length of the time
evolution for the GPE: $t=10^{4}\, T_{x}$, with $T_{x}=2\pi/\omega_{x}$.}
\end{figure}

In Fig. \ref{fig:frequency_shift_for_small_amplitudes} we compare
the results for the three different models: \textbf{(a)} the numerical
simulation, \textbf{(b)} the column density model, and \textbf{(c)}
the three dimensional Thomas-Fermi model. The numerical solution of
the GPE yields a discrete set of data points for the center-of-mass
position at different times $t_{n}$. From this we extract the
oscillation frequency with the help of a discrete Fourier transformation.
For the numerical calculations we use two different oscillation amplitudes,
$x_{s}=0.001\,\lambda_{x}^{(0)}$ and $x_{s}=0.1\,\lambda_{x}^{(0)}$.
Note, that we measure the oscillation amplitude $x_{s}$, as well
as the distance to the surface $x_{d}$, in units of $\lambda_{x}^{(0)}$.
It is the semi-axis of the Thomas-Fermi density distribution without
dipole-dipole interaction, i.e. for $\varepsilon_{D}=0$. If we used
the actual semi-axis $\lambda_{x}$ instead, the distance $x_{d}$
and amplitude $x_{s}$ would depend on $\varepsilon_{D}$. For the
results, presented in Fig. \ref{fig:frequency_shift_for_small_amplitudes}
and Fig. \ref{fig:frequency-shift-for-large-amplitudes}, we used
$\lambda_{x}^{(0)}=\unit{7}{\micro\meter}$. With that the distance
between the BEC and the surface is $x_{d}=\unit{14}{\micro\meter}$.
Experiments \cite{Kasch,Bernon}, as well as theoretical calculations
\cite{Markowsky,Dikovsky} have shown that such distances are realistic.
In Fig. \ref{fig:frequency_shift_for_small_amplitudes} we present
the frequency shift as a function of the trap aspect ratio $\nu=\omega_{x}/\omega_{z}$.
The other trap aspect ratio $\kappa=\omega_{y}/\omega_{x}$ remains
constant. Here we consider a cylindrical symmetric trap with $\kappa=1$,
which means that $\lambda_{x}^{(0)}=\lambda_{y}^{(0)}$. Since the
magnetic dipoles are oriented in the $z$-direction, we also have
$\lambda_{x}=\lambda_{y}$. The central density $n_{0}^{(0)}$ and
the semi-axes $\lambda_{x}^{(0)}$ shall remain constant for all values
of $\nu$. This is achieved by adjusting the number of atoms in the
BEC according to the aspect ratio $\nu$. If $\nu$ increases, also
the number of atoms must increase. The connection between $\nu$ and
the number of atoms can be established via the expression for the
central density. From this we find $\nu=\lambda_{z}^{(0)}/\lambda_{x}^{(0)}=\frac{15}{8\pi}\frac{N}{n_{0}^{(0)}\left[\lambda_{x}^{(0)}\right]^{2}\lambda_{y}^{(0)}}$.
In the numerical calculations for the frequency shift we set the central
density to be $n_{0}^{(0)}=\unit{2.5\times10^{13}}{\rpcubic{\centi\meter}}$,
where $n_{0}^{(0)}$ is the central density for the case $\varepsilon_{D}=0$.
The actual central density $n_{0}$ is somewhat modified due to the
dipole-dipole interaction. Fig. \ref{fig:frequency_shift_for_small_amplitudes}
shows that in the amplitude range from $x_{s}=0.001\,\lambda_{x}^{(0)}$
to $x_{s}=0.1\,\lambda_{x}^{(0)}$ the frequency shift does not change.
We also find an excellent agreement between the numerical results
and the results obtained using the 3D Thomas-Fermi model. Furthermore
we can see that the results obtained from the column density model
also show a good agreement with the numerical data. The largest deviations
can be found for smaller values of $\nu$, around the position of
the maximum frequency shift. For large aspect ratios of the trap the
results of all three models converge. If the aspect ratio is large,
the BEC is very elongated: $\lambda_{x},\lambda_{y}\ll\lambda_{z}$.
The more elongated the BEC, the better it can be approximated by a
one dimensional column density distribution. In the region where $\lambda_{x},\lambda_{y}\simeq\lambda_{z}$
the column density model is not a very good approximation. Of course,
the accuracy of the column density model also depends on the distance
to the surface. However, as we can see the model yields good results
for a distance of $x_{d}=2\lambda_{x}^{(0)}$. For larger distances
the accuracy increases. Since smaller distances are likely to be difficult
to achieve in experiment, we are not considering them here.

\subsubsection{Large amplitude oscillations}

\begin{figure}
\includegraphics[width=1\columnwidth]{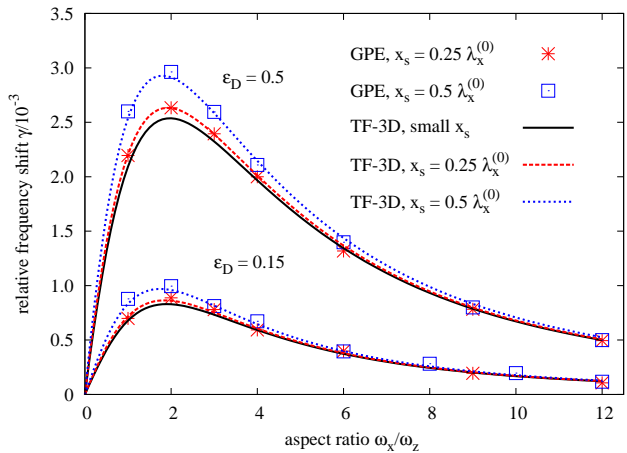}

\caption{\label{fig:frequency-shift-for-large-amplitudes}(Color online) Frequency shift for
large amplitude oscillations: Again the frequency shift is shown for
two different dipole-dipole interaction strengths, $\varepsilon_{D}=\varepsilon_{D}^{(m)}=0.15$
and $\varepsilon_{D}=\varepsilon_{D}^{(m)}=0.5$. The data points
show the frequency shift based on the numerical solution of the time
dependent GPE (\ref{eq:time_dep_GPE}), where two different oscillation
amplitudes are presented, $x_{s}=0.25\,\lambda_{x}^{(0)}$ (red stars)
and $x_{s}=0.5\,\lambda_{x}^{(0)}$ (blue squares). The lines show
the frequency shift based on the three dimensional Thomas-Fermi approximation.
The solid black line shows the result for small amplitude oscillations
(\ref{eq:gamma_3D_Thomas-Fermi}). The dashed red line and the dotted
blue line show the frequency shift with large amplitude correction
(\ref{eq:gamma_3D_Thomas-Fermi_large_amplitude}).\emph{ Other parameters}:
$\kappa=\omega_{y}/\omega_{x}=1$, length of the time evolution for
the GPE: $t=10^{4}\, T_{x}$, with $T_{x}=2\pi/\omega_{x}$.}

\end{figure}

So far we have only discussed small amplitude oscillations and have
seen that the resulting frequency shift can be described very accurately
using approximation (\ref{eq:gamma_3D_Thomas-Fermi}). In Fig. \ref{fig:frequency-shift-for-large-amplitudes}
we show the results for the frequency shift obtained from numerical
calculations with amplitude $x_{s}=0.5\,\lambda_{x}^{(0)}$. Here
we clearly see deviations from approximation (\ref{eq:gamma_3D_Thomas-Fermi}).
In order to describe this, we need to consider higher order corrections
to the frequency shift. Again, we follow the work of Antezza et al.
\cite{Antezza2004}, and the detailed calculation can be found in
Appendix \ref{sec:frequency_shift}. The resulting expression for
the frequency shift reads
\begin{equation}
\begin{split}\gamma=\frac{\omega_{x}^{\prime}-\omega_{x}}{\omega_{x}}=\frac{1}{2\omega_{x}^{2}M}\frac{1}{N}\int\mathrm{d}\mathbf{r}\, & n_{\mathrm{TF}}\left(\mathbf{r}\right)\Bigg[g\left(\mathbf{r};x_{d}\right)\\
 & +\frac{x_{s}^{2}}{8}h\left(\mathbf{r};x_{d}\right)\Bigg],
\end{split}
\label{eq:gamma_3D_Thomas-Fermi_large_amplitude}
\end{equation}
 with
\[
h\left(\mathbf{r};x_{d}\right)=16\left.\frac{\partial^{4}}{\partial x^{\prime4}}V_{\mathrm{mir}}\left(\mathbf{r}^{\prime}\right)\right|_{\mathbf{r}^{\prime}=\mathbf{r}+2x_{d}\hat{\mathbf{e}}_{x}}.
\]
The results obtained from this approximation are presented in Fig.
\ref{fig:frequency-shift-for-large-amplitudes}. We calculate the
frequency shift for two different oscillation amplitudes, $x_{s}=0.25\,\lambda_{x}^{(0)}$
and $x_{s}=0.5\,\lambda_{x}^{(0)}$. As one can see from Fig. \ref{fig:frequency-shift-for-large-amplitudes},
the frequency shift increases for larger amplitudes. The results from
the numerical calculations show an excellent agreement with approximation
(\ref{eq:gamma_3D_Thomas-Fermi_large_amplitude}). For $x_{s}=0.25\,\lambda_{x}^{(0)}$,
the correction to the small amplitude case, is only minor. Whereas
for $x_{s}=0.5\,\lambda_{x}^{(0)}$, the correction becomes more significant.
In the region around the maximum, the correction to the small amplitudes
is more important. The more the aspect ratio $\nu$ of the trap is
increased, the more do the results for different amplitudes converge.

\subsection{Excitation of collective modes due to the BEC-mirror interaction}

The center-of-mass motion is not the only collective mode of a BEC
where the eddy current effect can be observed. In the following we
will focus on the so-called monopole-quadrupole modes \cite{Sapina}.
In a harmonic trap the density distribution of a BEC within TF approximation
is an ellipsoid. Monopole-quadrupole modes are fluctuations of the
density, where the form of the BEC always remains ellipsoidal. This
means that the semi-axes become time dependent. Modes of this type
can be excited, for example, by a sudden change of the trap frequencies.
In a harmonic trap the center-of-mass motion and the monopole-quadrupole
modes are decoupled. If the trap minimum is shifted, only the center-of-mass
oscillation is excited, while the shape fluctuations remain unaffected.
However, if the trapping potential is not purely harmonic, this is
no longer the case. The potential generated by the superconducting
surface creates an anharmonicity of the potential which leads to a
coupling of said modes. If one of the monopole-quadrupole mode frequencies
coincides with the center-of-mass oscillation frequency $\omega_{x}^{\prime}$,
or with an integer multiple of $\omega_{x}^{\prime}$, a resonant
excitation appears. In the vicinity of a resonance the strength of
the excitation is enhanced, which increases the chance to observe
the effect. In \cite{Sapina2013} we have discussed this coupling
within the framework of an effective anharmonic potential, which included
a fourth order term of the form $x^{2}z^{2}$. This term generates
a coupling between the center-of-mass motion and the breather mode
of the BEC and a resonance occurs when the breather mode frequency
matches twice the center-of-mass oscillation frequency. As a measure
for the shape fluctuations we observe how the aspect ratio $a\left(t\right)=\sigma_{z}\left(t\right)/\sigma_{x}\left(t\right)$
of the BEC changes as a function of time. A discrete Fourier analysis
of this data yields information on the strength of the excitation
as a function of frequency $\Omega$.

\begin{figure*}
\includegraphics[width=1\textwidth]{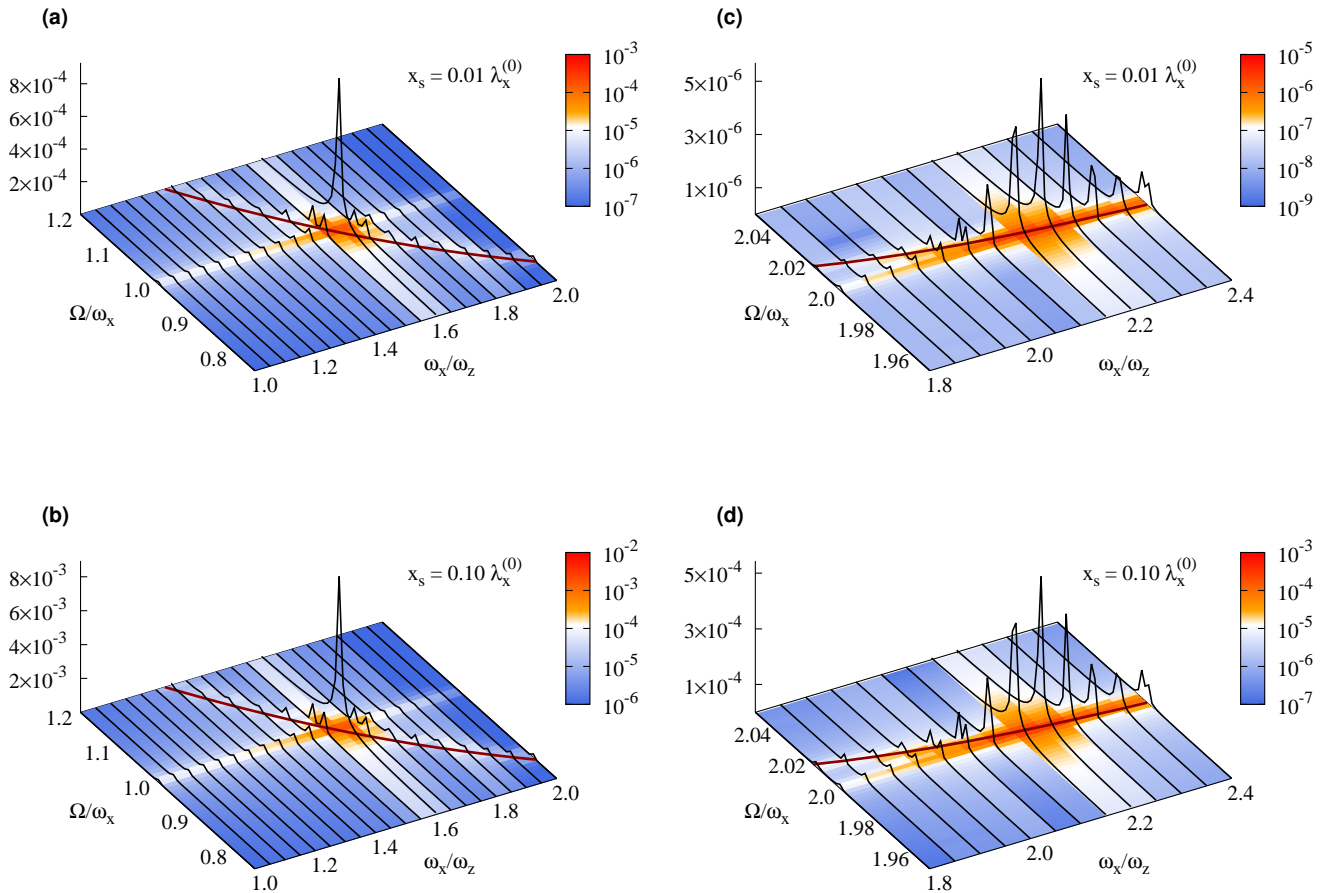}
\caption{\label{fig:sigma-x_spectrum_numerical_results_gpe}(Color online) Frequency spectra
for the relative fluctuation of the BEC aspect ratio $\Delta a\left(t\right)=\left(a\left(0\right)-a\left(t\right)\right)/a\left(0\right)$
of the BEC. The simulations were performed for various trap aspect
ratios $\nu=\omega_{x}/\omega_{z}$, ranging from $\nu=1$ to $\nu=2.4$
in steps of $\Delta\nu=0.05$. The plots in\textbf{ (a)} and \textbf{(b)}
show the region of the crossing point between the single oscillation
frequency $\omega_{x}^{\prime}$ and one of the monopole-quadrupole
modes. From\textbf{ (a)} to \textbf{(b)} the oscillation amplitude
increases by an order of magnitude, from $0.01\,\lambda_{x}^{(0)}$
to $0.1\,\lambda_{x}^{(0)}$. The resonance peak at the crossing also
increases by an order of magnitude. In \textbf{(c)} and \textbf{(d)}
the region of the crossing between the double oscillation frequency
$2\omega_{x}^{\prime}$ and the breather mode is presented. Again,
the amplitude $x_{s}$ increases from \textbf{(c)} to \textbf{(d)
}by an order of magnitude. The resonance peak increases by two orders
of magnitude. \emph{Parameters}: $\varepsilon_{D}=\varepsilon_{D}^{(m)}=0.2$;
$\kappa=\omega_{y}/\omega_{x}=0.99$; length of time evolution for
\textbf{(a)} and \textbf{(b)} $t=100\, T_{x}$ and for \textbf{(c)}
and \textbf{(d)} $t=500\, T_{x}$, with $T_{x}=2\pi/\omega_{x}$.}
\end{figure*}

In Fig. \ref{fig:sigma-x_spectrum_numerical_results_gpe} we present
the frequency spectrum of the BEC aspect ratio $a\left(t\right)$,
obtained from the numerical solution of GPE (\ref{eq:time_dep_GPE}).
We set the dipole-dipole interaction strength to be $\varepsilon_{D}=\varepsilon_{D}^{(m)}=0.2$.
We calculated the monopole-quadrupole mode frequencies within the
Thomas-Fermi approximation and indicate them in the plots as red lines
on the bottom. The simulations were performed for various trap aspect
ratios $\nu=\omega_{x}/\omega_{z}$, ranging from $\nu=1.0$ to $\nu=2.4$
in steps of $\Delta\nu=0.05$. The second trap aspect ratio $\kappa=\omega_{y}/\omega_{x}$
is set to $\kappa=0.99$. The spectrum for every aspect ratio is plotted
as a black line. The bottom color map shows the excitation on a logarithmic
scale, where blue indicates a weak excitation and red a strong excitation.
We compare two different oscillation amplitudes, $x_{s}=0.01\,\lambda_{x}^{(0)}$
and $x_{s}=0.1\,\lambda_{x}^{(0)}$. We set $\lambda_{x}^{(0)}$ to
be $\unit{7}{\micro\meter}$ and the central density is $n_{0}^{(0)}=\unit{5\times10^{13}}{\rpcubic{\centi\meter}}$.
For the $s$-wave scattering length we used the value for chromium,
which is $\unit{5.1}{\nano\meter}$ \cite{Chrom2}. Theses parameters
stay the same in every calculation, so that every aspect ratio corresponds
to a certain number of atoms in the BEC. One can see that the peaks
in the spectra compare quite nicely to the Thomas-Fermi mode frequencies.
This means the number of atoms is large enough such that we are within,
or at least close to the Thomas-Fermi regime. Fig. \ref{fig:sigma-x_spectrum_numerical_results_gpe}
\textbf{(c)} and \textbf{(d)} show the section of the spectrum where
the breather mode is located. As expected from our previous calculations
with the effective potential, we see a resonance at the position where
the breather mode frequency and the double oscillation frequency $2\omega_{x}^{\prime}$
coincide. If we used an symmetric trap with $\kappa=1$, the breather
mode frequency would approach the double oscillation frequency rather
than cross it \cite{Sapina}. The strength of the resonance depends
of course on the strength of the dipole-dipole interaction parameter
$\varepsilon_{D}^{(m)}$ and also on the amplitude $x_{s}$ of the
center-of-mass oscillation. In Fig. \ref{fig:sigma-x_spectrum_numerical_results_gpe}
\textbf{(c)} the oscillation amplitude is $x_{s}=0.01\,\lambda_{x}^{(0)}$
and in Fig. \ref{fig:sigma-x_spectrum_numerical_results_gpe} \textbf{(d)}
it is $x_{s}=0.1\,\lambda_{x}^{(0)}$. While we increase the oscillation
amplitude by one order of magnitude, the strength of the resonance
increases by two orders of magnitude. This suggests a quadratic dependence
of the resonance strength on the oscillation amplitude. 

In Fig. \ref{fig:sigma-x_spectrum_numerical_results_gpe} \textbf{(a)}
and \textbf{(b) }we present a different section of the spectrum. In
this section we find the lowest lying monopole-quadrupole mode. Again
we find a resonance peak in the spectrum, only this time the resonance
occurs at the position where the mode frequency crosses the single
oscillation frequency $\omega_{x}^{\prime}$. Again, the oscillation
amplitude from Fig. \ref{fig:sigma-x_spectrum_numerical_results_gpe}
\textbf{(a)} to \textbf{(b)} increases by one order of magnitude.
This time, also the strength of the resonance increases by one order
of magnitude. From this we can infer that this resonance peak grows
linearly with the oscillation amplitude $x_{s}$. 

To obtain a better understanding of the excitation mechanism, 
let us simplify the situation as follows.
The mirror potential generates an anharmonic perturbation to
the harmonic trapping potential. Expanding the mirror potential 
in a Taylor series yields the involved anharmonic terms.
Let us now consider the situation in the rest frame of the center-of-mass. 
In this frame the anharmonic terms of the potential lead to a 
time dependent curvature of the potential \cite{OttJPhysB,OttPRL}.
For example, if we transform the term $xz^{2}$ into the
rest frame, via $x=x^{\prime}+x_{s}\sin\left(\omega_{x}^{\prime}t\right)$,
the curvature in the $z^\prime$-direction gets a time dependent component:
$x_{s}\sin(\omega_{x}^{\prime}t)\cdot z^{\prime2}$. 
Obviously, this modulates the curvature of the potential in the 
rest frame with center-of-mass oscillation frequency $\omega_x^\prime$. 
A time dependent curvature leads to the excitation of collective modes \cite{Pitaevskii}.
If one of the modes happens to have the same frequency as the driving frequency, a resonance occurs. 
This picture also explains the scaling of the peak height with the oscillation amplitude.
The time dependent component which generates the resonance peak at $\Omega=\omega_x^\prime$
is linear in $x_s$. In contrast, the term $x^2z^2$ would create a modulation
of the form $x_{s}^2\sin^2(\omega_{x}^{\prime}t)$, which drives modes with
double oscillation frequency and is quadratic in $x_s$. This qualitatively explains the 
scaling of the resonance peak at $\Omega=2\omega_x^\prime$.

\section{Frequency shift for a different polarization of the Bose-Einstein condensate}

\begin{figure*}
\includegraphics[width=0.8\textwidth]{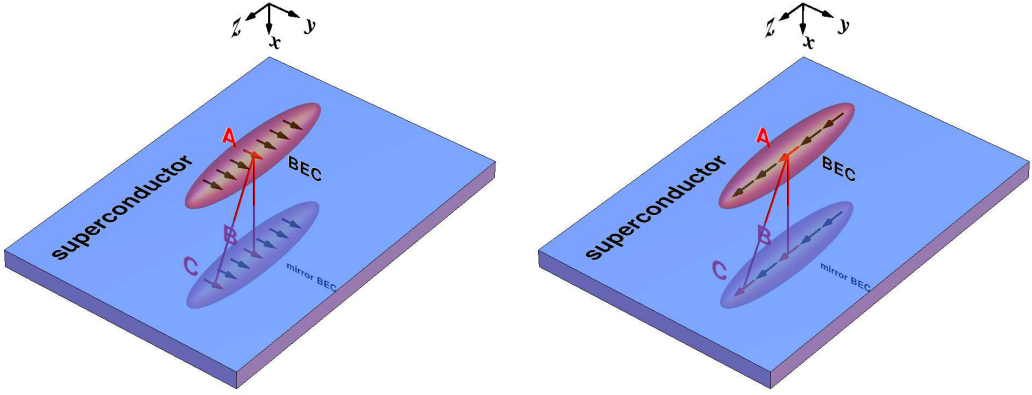}
\caption{\label{fig:setup_dipoles_oriented_in_z_direction}(Color online) The left setup depicts
the configuration where the dipoles are oriented in the $y$-direction.
The interaction between the dipoles $A$ and $B$ differs from the
interaction between $A$ and $C$ only in the distance. The relative
orientation of the dipoles is the same, therefore also the interaction
sign is the same. In this configuration all contributions add up constructively.
The right setup depicts the situation where all dipoles are oriented
in the $z$-direction. The relative orientation between $A$ and $B$
is different than the relative orientation between $A$ and $C$,
therefore also the interaction sign may change. In this configuration
the contributions from the edges partially compensate the contributions
from the center.}
\end{figure*}

In this section we discuss the dependence of the frequency shift on
the orientation of the dipoles. So far we have only considered the
case where the dipoles are oriented in the $z$-direction, which is
parallel to the superconductor surface and coincides with the long
axis of the BEC (see Fig. \ref{fig:BEC_and_mirror}). In principle,
the dipoles can be oriented in any direction in which an external
polarizing $\mathbf{B}$-field can be applied. Since the polarization
of the dipoles perpendicular to the surface might be difficult to
achieve in an experiment, we will not discuss this case here. However,
reorienting the polarization parallel to the surface should not pose
a problem. Let us assume that the dipoles are oriented in the $y$-direction,
while the direction of the long axis of the BEC remains the $z$-direction.
The $x$-direction is still perpendicular to the surface. Compared
to the setup we discussed earlier, the dipoles are now rotated by
$90^{\circ}$ parallel to the surface. The described setup is depicted
in Fig. \ref{fig:setup_dipoles_oriented_in_z_direction} (left panel). 

Let us compare the interaction between the dipoles $A$ and $B$ with
the interaction between the dipoles $A$ and $C$. The relative orientation
between the dipoles remains the same and only the distance changes.
This means that only the interaction strength is affected and not
the interaction sign. From the center towards the edges of the BEC
the interaction strength decreases. Since the interaction sign remains
the same, however, all contributions add up constructively to the
overall interaction. If we now increase the aspect ratio $\nu$ and
add more and more atoms (in such a way that the central density remains
the same), then we expect to see an increase of the frequency shift.
The frequency shift should increase monotonically with the number
of atoms in the BEC.

The situation is different if the dipoles are oriented in the $z$-direction.
In this case the interaction sign between $A$ and $B$ is not the
same as between $A$ and $C$. Contributions along the $z$-axis of
the BEC can cancel each other out. In this case, depending on the
length of the BEC, the overall interaction can be smaller than in
the case with the dipoles oriented in the $y$-direction. The longer
the BEC gets, the smaller is the overall interaction. In the limit
$\nu\rightarrow\infty$ the overall interaction, and also the frequency
shift, go to zero.

As we have already seen in section \ref{sub:Numerical_results_for_frequency_shift},
the column density model yields very good results, which is why we
will use it here to discuss the configuration with the dipoles oriented
in the $y$-direction. The potential along the axes of the BEC generated
by the dipoles of the mirror BEC is given by 
\begin{equation}
V_{\mathrm{mir}}\left(x,z\right)=\frac{g_{D}}{4\pi}\intop_{-\lambda_{z}}^{\lambda_{z}}\mathrm{d}z^{\prime}\,\frac{n_{1D}\left(z^{\prime}\right)}{\left[x^{2}+\left(z-z^{\prime}\right)^{2}\right]^{3/2}}.\label{eq:potential_generated_by_dipole_ey}
\end{equation}
From this expression we can already see, that the sign of $V_{\mathrm{mir}}$
does not change along the axis of the BEC. Given that the semi-axis
$\lambda_z$ is known, the expression can be evaluated completely analytically
and we obtain the resulting frequency shift with the method already discussed. 
Here we only give the necessary expressions to calculate the frequency shift, 
this model is explained in more detail in \cite{Sapina2013}. 
\begin{equation}
\gamma_{y}=\frac{\omega_{x}^{\prime}-\omega_{x}}{\omega_{x}}=\frac{1}{2M\omega_{x}^{2}}\frac{1}{N}\intop_{-\lambda_{z}}^{\lambda_{z}}\mathrm{d}z\, n_{1D}\left(z\right)g\left(z;x_{d}\right),\label{eq:frequency_shift_column_y-direction}
\end{equation}
with the column density
\[
n_{\mathrm{1D}}\left(z\right)=\frac{15}{16}\frac{N}{\lambda_{z}}\left(1-\frac{z^{2}}{\lambda_{z}^{2}}\right)^{2},
\]
and curvature change of the mirror interaction potential
\[
g\left(z;x_{d}\right)=4\left.\frac{\partial^{2}}{\partial x^{2}}V_{\mathrm{mir}}\left(x,z\right)\right|_{x=2x_{d}}.
\]
The factor $4$ in the curvature accounts for the fact that the motion
of the BEC leads also to motion of the mirror BEC. In order to calculate
the frequency shift one can determine the analytical expressions for
(\ref{eq:potential_generated_by_dipole_ey}) and then numerically
integrate (\ref{eq:frequency_shift_column_y-direction}). However,
the integral in (\ref{eq:frequency_shift_column_y-direction}) can
also be calculated completely analytical. The result is a very lengthy
expression, so we will not give it here but we will present an
interesting limit.

We will split the following discussion into two parts. In the first part we will
neglect the dipole-dipole interaction between the atoms, meaning we have 
$\varepsilon_D^{(m)}\neq0$ and $\varepsilon_D=0$. This approach is useful,
since it will provide exact analytical results for the frequency shift.
In the second part we will include the dipole-dipole interaction, i.e. 
$\varepsilon_D=\varepsilon_D^{(m)}\neq0$, and show that resulting corrections
are very small.

As we have already mentioned, $\gamma_{y}$ will
increase monotonically as a function of the trap aspect ratio $\nu=\omega_{x}/\omega_{z}$,
while the radial semi-axes and the central density are kept constant.
For the case $\varepsilon_D=0$, the Thomas-Fermi semi-axes are given
by a simple analytical expression (\ref{eq:TF-semi-axes-ed=0}), and we set $\lambda_a=\lambda_a^{(0)}$.
Using the analytical results for $\gamma_{y}$ and taking the limit
$\nu\rightarrow\infty$ we find
\begin{equation}
\gamma_{y}^{(\mathrm{max})}=\lim_{\nu\rightarrow\infty}\gamma_{y}=\frac{3}{14}\frac{\left[\lambda_{x}^{(0)}\right]^{4}}{x_{d}^{4}}\varepsilon_{D}^{(m)}.\label{eq:gamma-y-max}
\end{equation}
This expression only holds for $\varepsilon_{D}=0$. We will discuss
the corrections for $\varepsilon_{D}\neq0$ below. In \cite{Sapina2013}
we presented a similar value for the case that the dipoles are oriented
in the $z$-direction
\begin{equation}
\gamma_{z}^{(\mathrm{max})}=0.11\frac{\left[\lambda_{x}^{(0)}\right]^{4}}{x_{d}^{4}}\varepsilon_{D}^{(m)}.\label{eq:gamma-z-max}
\end{equation}
If we compare the two we see that $\gamma_{y}^{(\mathrm{max})}$ is
roughly by a factor of $2$ larger than $\gamma_{z}^{(\mathrm{max})}$.
By orienting the dipoles in the $y$-direction instead of the $z$-direction
the strength of the eddy current effect can be enhanced. The downside
is however, that $\gamma_{y}$ has not the same characteristic shape
as $\gamma_{z}$, where for an optimal length of the BEC a maximal
frequency shift can be observed. 

\begin{figure}
\includegraphics[width=0.4\columnwidth]{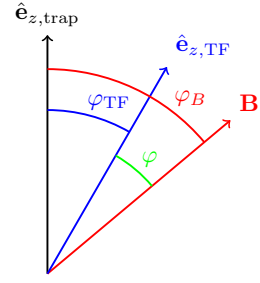}
\caption{\label{fig:angles}(Color online) The orientation of the Thomas-Fermi ellipsoid 
relative to the external polarizing field $\mathbf{B}$. 
The direction of the $z$-axis of the trap is indicated by $\hat{\mathbf{e}}_{z,\mathrm{trap}}$ 
and the direction of the BEC is indicated by $\hat{\mathbf{e}}_{z,\mathrm{TF}}$. 
Due to the dipole-dipole interaction those two are no longer aligned. 
This modifies the angle between the BEC axis and the magnetic field: $\varphi=\varphi_B-\varphi_\mathrm{TF}$.}
\end{figure}

Next, let us see what happens if the dipoles are oriented in an arbitrary
direction in the plane parallel to the surface. We will denote the
angle between the magnetic dipole moments and the long axis of the
Thomas-Fermi ellipsoid with $\varphi$ (see Fig. \ref{fig:angles}). That means we can write the
potential generated by the mirror BEC as 
\begin{eqnarray*}
V_{\mathrm{mir}}\left(x,z\right) & = & -\,\frac{g_{D}}{4\pi}\intop\mathrm{d}z^{\prime}\, n_{\mathrm{1D}}\left(z\right)\Bigg[\frac{3\left(z-z^{\prime}\right)^{2}\cos^{2}\varphi}{\left|\mathbf{r}-\mathbf{r}^{\prime}\right|^{5}}\\
 &  & -\frac{1}{\left|\mathbf{r}-\mathbf{r}^{\prime}\right|^{3}}\Bigg]\\
\\
 & = & -\,\frac{g_{D}}{4\pi}\intop\mathrm{d}z^{\prime}\, n_{\mathrm{1D}}\left(z\right)\Bigg\{\Bigg[\frac{3\left(z-z^{\prime}\right)^{2}}{\left|\mathbf{r}-\mathbf{r}^{\prime}\right|^{5}}\\
 &  & -\frac{1}{\left|\mathbf{r}-\mathbf{r}^{\prime}\right|^{3}}\Bigg]\cos^{2}\varphi-\frac{\sin^{2}\varphi}{\left|\mathbf{r}-\mathbf{r}^{\prime}\right|^{3}}\Bigg\}\\
\\
 & = & V_{\mathrm{mir}}^{(z)}\left(x,z\right)\cos^{2}\varphi+V_{\mathrm{mir}}^{(y)}\left(x,z\right)\sin^{2}\varphi.
\end{eqnarray*}
The interaction potential is merely a superposition of the two orientations
which have already been discussed. This means that also the frequency
shift can be constructed from the results we already know. We have
\[
\gamma\left(\varphi\right)=\gamma_{z}\cos^{2}\varphi+\gamma_{y}\sin^{2}\varphi,
\]
and for the case $\nu\rightarrow\infty$, we get
\begin{equation}
\gamma^{(\mathrm{max})}\left(\varphi\right)=\lim_{\nu\rightarrow\infty}\gamma\left(\varphi\right)=\frac{3}{14}\frac{\left[\lambda_{x}^{(0)}\right]^{4}}{x_{d}^{4}}\varepsilon_{D}^{(m)}\sin^{2}\varphi.\label{eq:gamma(phi)_large_nu}
\end{equation}

\begin{figure}
\includegraphics[width=1\columnwidth]{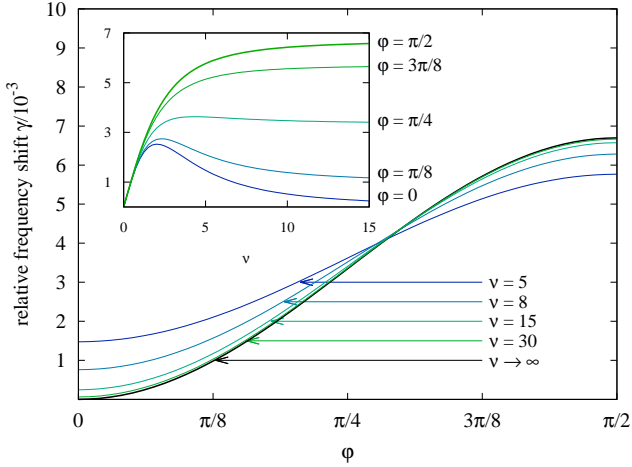}
\caption{\label{fig:gamma_vs_phi_ed=00003D0}(Color online) The frequency shift as a function
of the dipole orientation angle $\varphi$ for various trap aspect
ratios $\nu$. In the inset the frequency shift is shown as a function
of the trap aspect ratio $\nu$ for various orientation angles $\varphi$.
The curves were calculated with the column density model. \emph{Parameters}:
$\varepsilon_{D}^{(m)}=0.5$, $\varepsilon_{D}=0$, $\omega_{y}/\omega_{x}=1$,
and $x_{d}=2\lambda_{x}^{(0)}$.}
\end{figure}

In Fig. \ref{fig:gamma_vs_phi_ed=00003D0} we show the frequency shift
as a function of the orientation angle $\varphi$. One can see that
for the aspect ratio $\nu=15$ the limit (\ref{eq:gamma(phi)_large_nu})
is already a fairly good approximation. 

How does the dipole-dipole interaction between the atoms in the BEC itself influence
these results? 
For $\varepsilon_D\neq0$, there are two effects that need to be considered.
Firstly, the dipole-dipole interaction modifies the shape of the BEC.
Secondly, the orientation angel between the BEC and the dipoles changes.

\begin{figure}
\includegraphics[width=1\columnwidth]{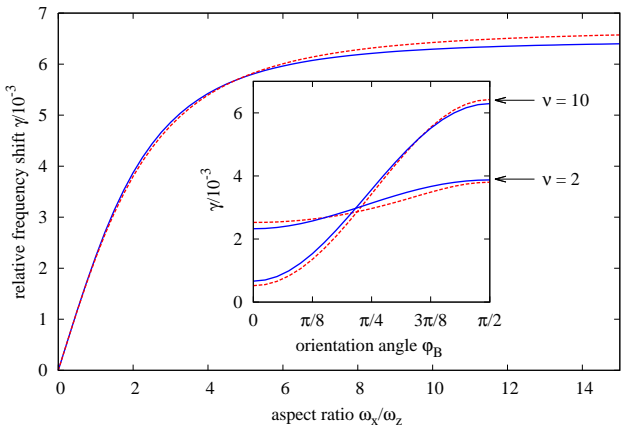}
\caption{\label{fig:gamma-y_vs_nu}(Color online) Frequency shift vs. aspect ratio for a BEC
with dipoles oriented in the $y$-direction. All curves were calculated
with the column density model. 
We compare the case $\varepsilon_D^{(m)}=0.5$ and $\varepsilon_D=0$ (red dashed lines)
to the case $\varepsilon_D^{(m)}=\varepsilon_D=0.5$ (blue solid lines).
In the inset the frequency shift is shown as a function of the magnetic field orientation 
angle $\varphi_B$ for two different values of $\nu$. Again, the two above mentioned cases are 
compared. Note that for $\varepsilon_D=0$ we have $\varphi=\varphi_B$, since $\varphi_\mathrm{TF}=0$.
}
\end{figure}

The change of the BEC shape has of course also an effect on the frequency shift. 
For the case that $\varepsilon_{D}\neq0$, we get an additional factor $\lambda_{z}^{(0)}/\lambda_{z}$ in the
expression for the frequency shift. 
In general, this factor needs to be calculated numerically.
For the dipoles oriented in the $y$-direction, the expression for the maximal frequency shift reads
\[
\gamma_{y}^{(\mathrm{max})}=\frac{3}{14}\frac{\left[\lambda_{x}^{(0)}\right]^{4}}{x_{d}^{4}}\varepsilon_{D}^{(m)}\lim_{\nu\rightarrow\infty}\frac{\lambda_{z}^{(0)}}{\lambda_{z}}.
\]
If we assume a cylindrical trap with $\omega_{x}=\omega_{y}>\omega_{z}$,
then magnetic repulsion between the atoms will cause the BEC to become
more elongated in the $z$-direction. Thus, we have $\lambda_{z}^{(0)}/\lambda_{z}<1$
and the limiting value for $\gamma_{y}^{(\mathrm{max})}$ is somewhat
smaller than given in (\ref{eq:gamma-y-max}). For a given dipole-dipole
interaction strength $\varepsilon_{D}$ the factor $\lim_{\nu\rightarrow\infty}\frac{\lambda_{z}^{(0)}}{\lambda_{z}}$
can be calculated. For $\varepsilon_{D}=0.1$ we find $\lim_{\nu\rightarrow\infty}\lambda_{z}^{(0)}/\lambda_{z}\approx0.99$
and for $\varepsilon_{D}=0.9$ we have $\lim_{\nu\rightarrow\infty}\lambda_{z}^{(0)}/\lambda_{z}\approx0.95$.
Even for large values of $\varepsilon_{D}$ the reduction of $\gamma_{y}^{(\mathrm{max})}$
is moderate. In Fig. \ref{fig:gamma-y_vs_nu} we show $\gamma_{y}$
for $\varepsilon_{D}=0$ as well as for $\varepsilon_{D}\neq0$. There
is only a minor difference between the two curves. 
This shows that expression (\ref{eq:gamma-y-max}) represents a very good
approximation for the maximally possible frequency shift.

Let us finally discuss the effect of $\varepsilon_D\neq0$ on the orientation angle.
Say the external polarizing field is oriented relative to the $z$-axis
of the trap in an angle $\varphi_{B}$. For the case that $0<\varphi_{B}<\pi/2$,
the resulting Thomas-Fermi ellipsoid is neither aligned with the magnetic
field nor with the harmonic trap.
The resulting configuration is depicted in Fig. \ref{fig:angles}.
This effect is discussed in more detail in Ref. \cite{Sapina}.
In order to calculate the frequency shift, we first need to determine the orientation angle
$\varphi_{\mathrm{TF}}$ of the BEC. This angle depends on $\varepsilon_{D}$,
the dipole orientation angle $\varphi_{B}$, and also on the geometry
of the trap. A set of self consistency equations is given in \cite{Sapina},
which can be used to determine the correct angle. Once we have $\varphi_{\mathrm{TF}}$,
we can also calculate $\varphi=\varphi_{B}-\varphi_{TF}$.
In the inset of Fig. \ref{fig:gamma-y_vs_nu} we show the frequency shift
as a function of $\varphi_B$ for two different trap aspect ratios. It is evident 
that the influence of $\varepsilon_D$ is only minor. Comparing the results
for $\nu=2$ to the results for $\nu=10$ shows that the influence of the 
dipole-dipole interaction becomes smaller for more elongated traps.
In the case of $\nu=2$ the maximal value for $\varphi_\mathrm{TF}$
is about $9\degree$, and for $\nu=10$ its value remains below $0.5\degree$.

The dependence of the frequency shift on the dipole orientation angle $\varphi$
is characteristic for the dipole-dipole interaction between the BEC
and its mirror. Therefore it is a fingerprint for the eddy current
effect which facilitates its experimental observation.

\section{Conclusion}

We have studied the effects of the magnetic interaction between a 
dipolar BEC and a superconductor on the dynamics of the BEC.  
The dynamical behaviour displays several features that can be used 
to identify and distinguish this effect from other effects
that might play a role close to the surface.
In particular we investigate the shift of the center-of-mass
oscillation frequency and also the excitation of BEC shape fluctuations.

The first characteristic is the change of the frequency shift with the number of atoms in the BEC.
We have discussed this already in Ref. \cite{Sapina2013}, where we used a relatively 
simple one dimensional model.
Here we use more sophisticated models, which show that the previously used model is not exact, 
but is a very useful tool to obtain analytical results which describes the qualitative 
behaviour and gives the correct order of magnitude for the effect.
Furthermore, we use the improved model to show how the frequency shift increases for large oscillation
amplitudes.

Another characteristic is the dependence of the frequency shift on the 
orientation of the magnetic dipoles of the atoms in the BEC.
To investigate this we used the one dimensional model from Ref. \cite{Sapina2013}.
In particular, we discussed the orientations of the dipoles parallel to the superconducting plane. 
Here, the characteristic dependence of the frequency shift on the orientation angle 
can be used as a fingerprint of the effect. We also showed that by
orienting the dipoles perpendicular to the long axis of the BEC, the effect can be increased 
by a factor of 2.

To investigate the excitation of collective modes of the BEC,
we use the results obtained from the numerical solution of the GPE.
The frequency spectrum shows two distinguished resonance peaks.
Each peak is connected to certain anharmonic terms in the potential
and shows a distinct scaling with the oscillation amplitude.
If the trap parameters are chosen properly, these two resonances
can significantly enhance the excitation of the collective modes.

In our calculations we assumed a distance of $\unit{14}{\micro\meter}$ 
between the superconducting surface and the minimum of the harmonic trap.
In Ref. \cite{Bernon} such a distance was demonstrated experimentally
in a superconducting microtrap. 
Theoretical calculations \cite{Markowsky,Dikovsky} suggest 
that even shorter distances are possible.

In the experiments thin superconducting strips or wires have been
used. In the present work we investigated a superconducting half space.
As has been discussed in Ref.~\cite{Sapina2013} a finite superconducting 
strip needs to meet certain requirements such that this approximation
is appropriate.
A strip thickness of twice the magnetic penetration depth is sufficient,
since the induced eddy currents only flow in the surface area
of the superconductor where the magnetic field penetrates
the superconductor.
The length and width of the strip should be larger than the BEC extensions
as well as the distance to the surface. If this is not the case, the effect
described here is reduced by a geometrical factor, which depends on the
solid angle under which the superconductor is seen by the BEC.

Until now only $^{87}$Rb BECs have been combined with superconductors.
Due to the small magnetic dipole moments of $^{87}$Rb the interaction
described here is rather small.
However, our results show that the combination of dipolar BECs with 
superconductors would open up the possibility to study this kind of interaction.

\begin{acknowledgments}
We acknowledge support by the DFG (SFB/TRR 21).
\end{acknowledgments}
\appendix

\section{Derivation of the GPE for a condensate interacting with its mirror\label{sec:appendix:mirror_term_GPE}}

Here we want to show how the GPE for a dipolar BEC close to a superconducting
surface can be derived. We will include an interaction term to the
many body Hamiltonian and then minimize the energy functional using
a Hartree \foreignlanguage{american}{ansatz} for the many body wave
function. Let us start with the Hamiltonian:
\begin{eqnarray*}
\hat{H} & = & \underbrace{\sum_{i=1}^{N}\left[\frac{\mathbf{p}_{i}^{2}}{2m}+V_{T}\left(\mathbf{r}_{i}\right)\right]+\frac{1}{2}\sum_{i=1}^{N}\sum_{j\neq i}^{N}U\left(\mathbf{r}_{i},\mathbf{r}_{j}\right)}_{\hat{H}_{0}}\\
 &  & +\underbrace{\sum_{i=1}^{N}\sum_{k=1}^{N}U_{\mathrm{md}}\left(\mathbf{r}_{i},\mathbf{r}_{k}^{\prime}\right)}_{\hat{H}_{\mathrm{mir}}}.
\end{eqnarray*}
The first part, denoted with $\hat{H}_{0}$, is the standard Hamiltonian.
It includes the kinetic term, the external potential, and interaction
between the particles in the BEC. The last part of the Hamiltonian,
denoted $\hat{H}_{\mathrm{mir}}$, describes the interaction between
the particles in the BEC and the mirror BEC. The index $i$ numbers
the atoms in the BEC and the index $k$ the atoms in the mirror BEC.
Since every atom can of course also interact with its mirror, we do
not need to make the restriction $i\neq k$. Also, every atom $i$
interacts with every mirror atom $k$, so that the factor $\frac{1}{2}$
is not needed. Minimizing the functional $E=\left\langle \Psi_{H}\right|\hat{H}_{0}\left|\Psi_{H}\right\rangle $,
under the constraint of particle conservation yields the Gross-Pitaevskii
equation (\ref{eq:stationary_GPE}). Let us now calculate the additional
term to this GPE, generated by $\hat{H}_{\mathrm{mir}}$. We make
the following Hartree ansatz
\[
\Phi_{H}\equiv\Phi_{H}\left(\mathbf{r}_{1},\ldots,\mathbf{r}_{N};\mathbf{r}_{1}^{\prime},\ldots,\mathbf{r}_{N}^{\prime}\right)=\prod_{i=1}^{N}\psi\left(\mathbf{r}_{i}\right)\prod_{k=1}^{N}\chi\left(\mathbf{r}_{k}^{\prime}\right),
\]
where $\psi\left(\mathbf{r}_{i}\right)$ are the single particle wave
functions of the atoms in the BEC and $\chi\left(\mathbf{r}_{k}^{\prime}\right)$
the single particle wave functions of the atoms in the mirror BEC.
Of course, a mirror atom does not have an actual wave function. However,
this picture is still valid, as long as the wave functions of the
atoms and the mirror atoms are well separated. For a better overview
we introduce the following abbreviations
\[
\mathrm{d}\mathbf{R}\equiv\mathrm{d}\mathbf{r}_{1}\ldots\mathrm{d}\mathbf{r}_{N},\quad\mathrm{d}\mathbf{R}^{\prime}\equiv\mathrm{d}\mathbf{r}_{1}^{\prime}\ldots\mathrm{d}\mathbf{r}_{N}^{\prime},
\]
\[
\psi_{i}\equiv\psi\left(\mathbf{r}_{i}\right),\quad\chi_{k}\equiv\chi\left(\mathbf{r}_{k}^{\prime}\right),\quad\text{and}\qquad U_{ik}\equiv U_{\mathrm{md}}\left(\mathbf{r}_{i},\mathbf{r}_{k}^{\prime}\right).
\]
Since the operators contained in $\hat{H}_{0}$ only act on atoms
in the BEC we have
\begin{eqnarray*}
\left\langle \Phi_{H}\right|\hat{H}_{0}\left|\Phi_{H}\right\rangle  & = & \intop\mathrm{d}\mathbf{R}\,\mathrm{d}\mathbf{R}^{\prime}\,\Phi_{H}^{*}\hat{H}_{0}\Phi_{H}\\
\\
 & = & \intop\mathrm{d}\mathbf{R}\,\mathrm{d}\mathbf{R}^{\prime}\,\prod_{i,k=1}^{N}\psi_{i}^{*}\chi_{k}^{*}\hat{H}_{0}\prod_{l,m}^{N}\psi_{l}\chi_{m}\\
\\
 & = & \intop\mathrm{d}\mathbf{R}\,\prod_{i,m=1}^{N}\psi_{i}^{*}\hat{H}_{0}\psi_{l}\underbrace{\prod_{k,m=1}^{N}\intop\mathrm{d}\mathbf{R}^{\prime}\chi_{k}^{*}\chi_{m}}_{=1}\\
\\
 & = & \left\langle \Psi_{H}\right|\hat{H}_{0}\left|\Psi_{H}\right\rangle ,
\end{eqnarray*}
with 
\[
\Psi_{H}\left(\mathbf{r}_{1},\ldots,\mathbf{r}_{N}\right)=\prod_{i=1}^{N}\psi\left(\mathbf{r}_{i}\right).
\]
>From that we get the usual GPE (\ref{eq:stationary_GPE}). Additional
terms to the GPE arise from $\hat{H}_{\mathrm{mir}}$. The energy
functional reads
\begin{eqnarray*}
\left\langle \hat{H}_{\mathrm{mir}}\right\rangle  & = & \left\langle \Phi_{H}\right|\hat{H}_{\mathrm{mir}}\left|\Phi_{H}\right\rangle \\
\\
 & = & \intop\mathrm{d}\mathbf{R}\,\mathrm{d}\mathbf{R}^{\prime}\,\prod_{i,k=1}^{N}\psi_{i}^{*}\chi_{k}^{*}\sum_{p=1}^{N}\sum_{q=1}^{N}U_{pq}\prod_{l,m}^{N}\psi_{l}\chi_{m}\\
\\
 & = & \sum_{p=1}^{N}\sum_{q=1}^{N}\intop\mathrm{d}\mathbf{r}_{p}\,\mathrm{d}\mathbf{r}_{q}^{\prime}\,\psi_{p}^{*}\chi_{q}^{*}U_{pq}\psi_{p}\chi_{q}\\
\\
 & = & N^{2}\intop\mathrm{d}\mathbf{r}\,\mathrm{d}\mathbf{r}^{\prime}\,\psi^{*}\chi^{*}U_{\mathrm{md}}\left(\mathbf{r},\mathbf{r}^{\prime}\right)\psi\chi.
\end{eqnarray*}
For the variational calculation we need to calculate the term $\frac{\partial}{\partial\psi^{*}}\left\langle \hat{H}_{\mathrm{mir}}\right\rangle $.
By doing this, we have to keep in mind that $\psi^{*}$ and $\chi^{*}$
are not independent functions. $\chi^{*}$ is the mirror function
of $\psi^{*}$, they are connected via
\[
\chi\left(x,y,z\right)=\psi\left(-x+2x_{d},y,z\right).
\]
We find
\begin{eqnarray*}
\frac{\partial}{\partial\psi^{*}}\left\langle \hat{H}_{\mathrm{mir}}\right\rangle  & = & N^{2}\intop\mathrm{d}\mathbf{r}\,\mathrm{d}\mathbf{r}^{\prime}\,\chi^{*}U_{\mathrm{md}}\left(\mathbf{r},\mathbf{r}^{\prime}\right)\psi\chi\\
 &  & +N^{2}\intop\mathrm{d}\mathbf{r}\,\mathrm{d}\mathbf{r}^{\prime}\,\psi^{*}U_{\mathrm{md}}\left(\mathbf{r},\mathbf{r}^{\prime}\right)\psi\chi\\
\\
 & = & 2N^{2}\intop\mathrm{d}\mathbf{r}\,\mathrm{d}\mathbf{r}^{\prime}\, U_{\mathrm{md}}\left(\mathbf{r},\mathbf{r}^{\prime}\right)\left|\chi\left(\mathbf{r}^{\prime}\right)\right|^{2}\psi\left(\mathbf{r}\right),
\end{eqnarray*}
the last line can be obtained by substituting in the second term $\tilde{x}=-x+2x_{d}$
and using that $U_{\mathrm{md}}\left(\mathbf{r},\mathbf{r}^{\prime}\right)=U_{\mathrm{md}}\left(\mathbf{r}^{\prime},\mathbf{r}\right)$.
>From the minimization of $E-\mu N$ we now find the stationary GPE
\begin{eqnarray*}
\mu\psi\left(\mathbf{r}\right) & = & \Bigg(-\frac{\hbar^{2}}{2M}\boldsymbol{\nabla}^{2}+V\left(\mathbf{r}\right)+N\int\mathrm{d\mathbf{r}}^{\prime}\, U\left(\mathbf{r},\mathbf{r}^{\prime}\right)\left|\psi\left(\mathbf{r}^{\prime}\right)\right|^{2}\\
\\
 &  & +2N\intop\mathrm{d}\mathbf{r}^{\prime}\, U_{\mathrm{md}}\left(\mathbf{r},\mathbf{r}^{\prime}\right)\left|\chi\left(\mathbf{r}^{\prime}\right)\right|^{2}\Bigg)\psi\left(\mathbf{r}\right).
\end{eqnarray*}

\section{\label{sec:appendix:calculation_mirror_term}Calculating the mirror term}

Here we calculate the mirror term generated by the mirror BEC. We
assume that the density distribution of the mirror has an ellipsoidal
shape. We need to determine the second derivative of the potential
function $\phi\left(\mathbf{r}\right)$ given in (\ref{eq:chandrasekhar_integral_1D}).
For the sake of convenience let us define the following function:
\[
\alpha\left(\mathbf{r},\, u\right)=\frac{\left(1-\frac{x^{2}}{\lambda_{x}^{2}+u}-\frac{y^{2}}{\lambda_{y}^{2}+u}-\frac{z^{2}}{\lambda_{z}^{2}+u}\right)^{2}}{\sqrt{\beta\left(u\right)}},
\]
with $\beta\left(u\right)=\left(\lambda_{x}^{2}+u\right)\left(\lambda_{y}^{2}+u\right)\left(\lambda_{z}^{2}+u\right)$,
so that we have
\[
\phi\left(\mathbf{r}\right)=\frac{\lambda_{x}\lambda_{y}\lambda_{z}}{8}\intop_{W\left(\mathbf{r}\right)}^{\infty}\mathrm{d}u\,\alpha\left(\mathbf{r},\, u\right).
\]
The first derivative with respect to $z$ reads 
\begin{eqnarray*}
\frac{\partial}{\partial z}\phi\left(\mathbf{r}\right) & = & \frac{\lambda_{x}\lambda_{y}\lambda_{z}}{8}\frac{\partial}{\partial z}\intop_{W\left(\mathbf{r}\right)}^{\infty}\mathrm{d}u\,\alpha\left(\mathbf{r},\, u\right)\\
\\
 & = & \frac{\lambda_{x}\lambda_{y}\lambda_{z}}{8}\Bigg[-\alpha\left(\mathbf{r},\, W\left(\mathbf{r}\right)\right)\frac{\partial W\left(\mathbf{r}\right)}{\partial z}\\
 &  & +\intop_{W\left(\mathbf{r}\right)}^{\infty}\mathrm{d}u\,\frac{\partial}{\partial z}\alpha\left(\mathbf{r},\, u\right)\Bigg].
\end{eqnarray*}
We can say from (\ref{eq:ellipsoidal_coordinate_W}) that $\alpha\left(\mathbf{r},\, W\left(\mathbf{r}\right)\right)=0$,
which leads to 
\[
\frac{\partial}{\partial z}\phi\left(\mathbf{r}\right)=\frac{\lambda_{x}\lambda_{y}\lambda_{z}}{8}\intop_{W\left(\mathbf{r}\right)}^{\infty}\mathrm{d}u\,\frac{\partial}{\partial z}\alpha\left(\mathbf{r},\, u\right).
\]
With that we find for the second derivative 
\begin{eqnarray*}
\frac{\partial^{2}}{\partial z^{2}}\phi\left(\mathbf{r}\right) & = & \frac{\lambda_{x}\lambda_{y}\lambda_{z}}{8}\frac{\partial}{\partial z}\intop_{W\left(\mathbf{r}\right)}^{\infty}\mathrm{d}u\,\frac{\partial}{\partial z}\alpha\left(\mathbf{r},\, u\right)\\
\\
 & = & -\left.\frac{\partial}{\partial z}\alpha\left(\mathbf{r},\, u\right)\right|_{u=W\left(\mathbf{r}\right)}\frac{\partial W\left(\mathbf{r}\right)}{\partial z}\\
 &  & +\intop_{W\left(\mathbf{r}\right)}^{\infty}\mathrm{d}u\,\frac{\partial^{2}}{\partial z^{2}}\alpha\left(\mathbf{r},\, u\right).
\end{eqnarray*}
Let us next determine the derivatives of $\alpha\left(\mathbf{r},\, u\right)$,
the first derivative reads
\[
\frac{\partial}{\partial z}\alpha\left(\mathbf{r},\, u\right)=\frac{-4\frac{z}{\lambda_{z}^{2}+u}\left(1-\frac{x^{2}}{\lambda_{x}^{2}+u}-\frac{y^{2}}{\lambda_{y}^{2}+u}-\frac{z^{2}}{\lambda_{z}^{2}+u}\right)}{\sqrt{\beta\left(u\right)}},
\]
so that we find
\[
\left.\frac{\partial}{\partial r_{z}}\alpha\left(\mathbf{r},\, u\right)\right|_{u=W\left(\mathbf{r}\right)}=0,
\]
and we therefore get 
\[
\frac{\partial^{2}}{\partial z^{2}}\phi\left(\mathbf{r}\right)=\frac{\lambda_{x}\lambda_{y}\lambda_{z}}{8}\intop_{W\left(\mathbf{r}\right)}^{\infty}\mathrm{d}u\,\frac{\partial^{2}}{\partial z^{2}}\alpha\left(\mathbf{r},\, u\right).
\]
The second derivative of $\alpha\left(\mathbf{r},\, u\right)$ is
given by 
\[
\frac{\partial^{2}}{\partial z^{2}}\alpha\left(\mathbf{r},\, u\right)=4\frac{\frac{2z^{2}}{\left(\lambda_{z}^{2}+u\right)}-\left(1-\frac{x^{2}}{\lambda_{x}^{2}+u}-\frac{y^{2}}{\lambda_{y}^{2}+u}-\frac{z^{2}}{\lambda_{z}^{2}+u}\right)}{\left(\lambda_{z}^{2}+u\right)\sqrt{\beta\left(u\right)}},
\]
so that we obtain our final result
\begin{eqnarray*}
\frac{\partial^{2}\phi\left(\mathbf{r}\right)}{\partial z^{2}} & = & -\frac{\lambda_{x}\lambda_{y}\lambda_{z}}{2}\\
\\
 &  & \times\intop_{W\left(\mathbf{r}\right)}^{\infty}\mathrm{d}u\,\frac{\left(1-\frac{x^{2}}{\lambda_{x}^{2}+u}-\frac{y^{2}}{\lambda_{y}^{2}+u}-3\frac{z^{2}}{\lambda_{z}^{2}+u}\right)}{\left(\lambda_{z}^{2}+u\right)\sqrt{\beta\left(u\right)}}.
\end{eqnarray*}

\section{\label{sec:frequency_shift}Frequency shift of the center-of-mass motion}

Here we show in more detail how we derived the results for the frequency
shift. We follow the approach of Antezza et al. \cite{Antezza2004},
where they discussed the effect of the Casimir-Polder force on the
center-of-mass motion of a BEC. Say we have a BEC in a harmonic trap
of the form (\ref{eq:harmonic_potential}) and the atoms interact
with the surface via a potential $U\left(\mathbf{r}\right)$. The
BEC is oscillating in the $x$-direction, which is perpendicular to
the surface. Antezza et al. show that in such a case the motion of
the center-of-mass is described by the following differential equation
\begin{eqnarray*}
\frac{\mathrm{d}^{2}x_{\mathrm{c.m.}}\left(t\right)}{\mathrm{d}t^{2}} & = & -\omega_{x}^{2}\cdot x_{\mathrm{c.m.}}\left(t\right)\\
 &  & -\frac{1}{MN}\int d\mathbf{r}\, n\left(\mathbf{r}-x_{\mathrm{c.m}.}\left(t\right)\hat{\mathbf{e}}_{x}\right)\frac{\partial}{\partial x}U\left(\mathbf{r}\right),
\end{eqnarray*}
where $n$ is the density distribution of the BEC and $x_{\mathrm{c.m}.}$
is the $x$-coordinate of the center-of-mass. For the density distribution
of the BEC we use the Thomas-Fermi approximation, which yields the
density distribution given in (\ref{eq:TF-density}). We can now perform
the shift $\mathbf{r}^{\prime}=\mathbf{r}-x_{\mathrm{c.m.}}\left(t\right)\hat{\mathbf{e}}_{x}$
and then define the following time dependent function
\[
Q\left(t\right)\equiv\int d\mathbf{r}^{\prime}\, n_{\mathrm{TF}}\left(\mathbf{r}^{\prime}\right)\frac{\partial}{\partial x^{\prime}}U\left(\mathbf{r}^{\prime}+x_{\mathrm{c.m}.}\left(t\right)\hat{\mathbf{e}}_{x}\right).
\]
In the center-of-mass system of the BEC the surface potential appears
to be a time dependent potential. Next we expand $U\left(\mathbf{r}+x_{\mathrm{c.m.}}\left(t\right)\hat{\mathbf{e}}_{x}\right)$
in terms of $x_{\mathrm{c.m.}}$. For small amplitude oscillations
it is sufficient to linearize $U$. However, if one is interested
in corrections due to large amplitudes, the series expansion needs
to be performed at least to third order in $x_{\mathrm{c.m.}}$. The
series expansion of $U$ can now be inserted back into $Q\left(t\right)$.
We assume that the center-of-mass performs a harmonic oscillation
of the form $x_{\mathrm{c.m}.}\left(t\right)=x_{s}\cos\left(\omega_{x}^{\prime}t\right)$,
so that we can expand $Q\left(t\right)$ in a Fourier series $Q\left(t\right)=\frac{a_{0}}{2}+\sum_{n=1}^{\infty}a_{n}\cos\left(\omega_{x}^{\prime}n\cdot t\right)$.
Since we are only interested in the frequency shift, we only need
to evaluate the term proportional to $\cos\left(\omega_{x}^{\prime}t\right)$.
The Fourier coefficient of this term reads
\[
a_{1}=\int d\mathbf{r}\, n_{\mathrm{TF}}\left(\mathbf{r}\right)\left[x_{s}\frac{\partial^{2}}{\partial x^{2}}U\left(\mathbf{r}\right)+\frac{x_{s}^{3}}{8}\frac{\partial^{4}}{\partial x^{4}}U\left(\mathbf{r}\right)\right].
\]
Inserting everything back in the equation of motion for the center-of-mass,
the difference between the squares of the frequencies is found to
be
\[
\omega_{x}^{\prime2}-\omega_{x}^{2}=\frac{1}{MN}\int d\mathbf{r}\, n_{\mathrm{TF}}\left(\mathbf{r}\right)\left[\frac{\partial^{2}}{\partial x^{2}}U\left(\mathbf{r}\right)+\frac{x_{s}^{2}}{8}\frac{\partial^{4}}{\partial x^{4}}U\left(\mathbf{r}\right)\right].
\]
In our case the surface potential $U\left(\mathbf{r}\right)$ can
be understood as the dipole-dipole interaction potential between the
atoms in the BEC and the mirror BEC
\[
U\left(\mathbf{r}\right)\rightarrow V_{\mathrm{mir}}\left(\mathbf{r}\right)=\intop\mathrm{d}\mathbf{r}^{\prime}\, n\left(\mathbf{r}^{\prime}\right)U_{\mathrm{md}}\left(\mathbf{r},\mathbf{r}^{\prime}\right),
\]
where $n$ is the density distribution of the mirror BEC. Since we
use the Thomas-Fermi approximation, the potential $V_{\mathrm{mir}}$,
and its derivatives, are most conveniently calculated using the index
integrals. If the dipoles are oriented in the $z$-direction, the
potential $V_{\mathrm{mir}}$ is essentially given by $\frac{\partial^{2}}{\partial z^{2}}\phi\left(\mathbf{r}\right)$,
for which the expression is presented in (\ref{eq:ddz_phi_with_index_integrals}).
As the BEC oscillates perpendicular to the surface, the mirror BEC
oscillates as well. However, the mirror BEC oscillates in opposite
phase to the BEC. To compensate for this, the above derivatives with
respect to $x$ need to be replaced by derivatives with respect to
$x/2$:

\begin{eqnarray*}
\frac{\partial^{2}}{\partial x^{2}}U\left(\mathbf{r}\right) & \rightarrow & 4\left.\frac{\partial^{2}}{\partial x^{\prime2}}V_{\mathrm{mir}}\left(\mathbf{r}^{\prime}\right)\right|_{\mathbf{r}^{\prime}=\mathbf{r}+2x_{d}\hat{\mathbf{e}}_{x}}\\
\\
\frac{\partial^{4}}{\partial x^{4}}U\left(\mathbf{r}\right) & \rightarrow & 16\left.\frac{\partial^{4}}{\partial x^{\prime4}}V_{\mathrm{mir}}\left(\mathbf{r}^{\prime}\right)\right|_{\mathbf{r}^{\prime}=\mathbf{r}+2x_{d}\hat{\mathbf{e}}_{x}}.
\end{eqnarray*}
The reason, that we evaluate the derivatives at the position $\mathbf{r}^{\prime}=\mathbf{r}+2x_{d}\hat{\mathbf{e}}_{x}$,
is simply the fact that we calculate $V_{\mathrm{mir}}\left(\mathbf{r}^{\prime}\right)$
in the frame of reference where the center of the mirror BEC is at
the origin. We will not give the expressions of the derivatives of
$V_{\mathrm{mir}}$ here, since they are rather long. Finally, we
can write $\omega_{x}^{\prime2}-\omega_{x}^{2}=\left(\omega_{x}^{\prime}-\omega_{x}\right)\left(\omega_{x}^{\prime}+\omega_{x}\right)\approx\left(\omega_{x}^{\prime}-\omega_{x}\right)2\omega_{x}$,
in the case that the difference between the harmonic trap frequency
$\omega_{x}$ and the frequency of the center-of-mass motion $\omega_{x}^{\prime}$
is small. In the case of small amplitudes we can neglect the term
quadratic in $x_{s}$, which then yields result (\ref{eq:gamma_3D_Thomas-Fermi}).
If we also consider the correction term, we find result (\ref{eq:gamma_3D_Thomas-Fermi_large_amplitude})
for the frequency shift.

\section{\label{sec:appendix:the_index_integrals}The index integrals}

Integrals of the type
\[
F\left(x,y,z\right)=\intop_{0}^{\infty}\mathrm{d}u\,\frac{1}{\left(x+u\right)^{1/2}}\frac{1}{\left(y+u\right)^{1/2}}\frac{1}{\left(z+u\right)^{3/2}},
\]
can be calculated numerically using the Carlson method \cite{Carlson}.
The algorithm for this is provided in \cite{NumericalRecipes}. With
that the index integrals
\[
I_{a}\left(\lambda_{x}^{2},\lambda_{y}^{2},\lambda_{z}^{2}\right)=\intop_{0}^{\infty}\mathrm{d}u\,\frac{1}{\sqrt{\beta\left(u\right)}}\frac{1}{\left(\lambda_{a}^{2}+u\right)},\quad a\in\left\{ x,y,z\right\} 
\]
can be calculated via 
\begin{eqnarray*}
I_{x}\left(\lambda_{x}^{2},\lambda_{y}^{2},\lambda_{z}^{2}\right) & = & F\left(\lambda_{y}^{2},\lambda_{z}^{2},\lambda_{x}^{2}\right),\\
\\
I_{y}\left(\lambda_{x}^{2},\lambda_{y}^{2},\lambda_{z}^{2}\right) & = & F\left(\lambda_{z}^{2},\lambda_{x}^{2},\lambda_{y}^{2}\right),\\
\\
I_{z}\left(\lambda_{x}^{2},\lambda_{y}^{2},\lambda_{z}^{2}\right) & = & F\left(\lambda_{x}^{2},\lambda_{y}^{2},\lambda_{z}^{2}\right),
\end{eqnarray*}
where the different index integrals have been constructed by a permutation
of the arguments of $F$. Actually, we do not need to calculate all
three integrals, since there exists a sum rule, which reads
\[
I_{x}+I_{y}+I_{z}=\frac{2}{\lambda_{x}\lambda_{y}\lambda_{z}}.
\]
With that it suffices to calculate only two of the three integrals.
In the case that the ellipsoid is uni-axial with $\lambda_{x}=\lambda_{y}$
the solution of these integrals can be given in a closed analytic
form:
\[
I_{x}=I_{y}=-\frac{\frac{\lambda_{z}}{\lambda_{x}}\sqrt{1-\frac{\lambda_{z}^{2}}{\lambda_{x}^{2}}}+\arcsin\left(\frac{\lambda_{z}}{\lambda_{x}}\right)-\frac{\pi}{2}}{\left(\lambda_{x}^{2}-\lambda_{z}^{2}\right)^{3/2}},
\]
\[
I_{z}=2\frac{\sqrt{\frac{\lambda_{x}^{2}}{\lambda_{z}^{2}}-1}+\arcsin\left(\frac{\lambda_{z}}{\lambda_{x}}\right)-\frac{\pi}{2}}{\left(\lambda_{x}^{2}-\lambda_{z}^{2}\right)^{3/2}}.
\]
Again, the sum rule can be used in order to calculate only one of
the two integrals. For a spherical BEC with $\lambda_{x}=\lambda_{y}=\lambda_{z}$
we get
\[
I_{x}=I_{y}=I_{z}=\frac{2}{3}\frac{1}{\lambda_{x}^{3}},
\]
which can be easily seen from the sum rule. In order to calculate
the mirror potential we need the index integrals 
\[
J_{a}\left(\lambda_{x}^{2},\lambda_{y}^{2},\lambda_{z}^{2}\right)=\intop_{W}^{\infty}\mathrm{d}u\,\frac{1}{\sqrt{\beta\left(u\right)}}\frac{1}{\left(\lambda_{a}^{2}+u\right)},\quad a\in\left\{ x,y,z\right\} ,
\]
instead of $I_{a}$. By substituting $\lambda_{a}^{2}\rightarrow\lambda_{a}^{2}+W$
we can obtain $J_{a}$ from $I_{a}$
\[
J_{a}\left(\lambda_{x}^{2},\lambda_{y}^{2},\lambda_{z}^{2}\right)=I_{a}\left(\lambda_{x}^{2}+W,\lambda_{y}^{2}+W,\lambda_{z}^{2}+W\right).
\]
For the integral $J_{a}$ the sum rule needs to be modified, it reads
\[
J_{x}+J_{y}+J_{z}=\frac{2}{\sqrt{\left(\lambda_{x}^{2}+W\right)\left(\lambda_{y}^{2}+W\right)\left(\lambda_{z}^{2}+W\right)}}.
\]
 From the single index integrals we now need to construct the double
index integrals
\[
J_{ab}\left(\lambda_{x}^{2},\lambda_{y}^{2},\lambda_{z}^{2}\right)=\intop_{W}^{\infty}\mathrm{d}u\,\frac{1}{\sqrt{\beta\left(u\right)}}\frac{1}{\left(\lambda_{a}^{2}+u\right)}\frac{1}{\left(\lambda_{b}^{2}+u\right)},
\]
with $a,b\in\left\{ x,y,z\right\} $. The two types of integrals are
connected via
\[
J_{ab}=-\frac{J_{a}-J_{b}}{\lambda_{a}^{2}-\lambda_{b}^{2}}.
\]
If we have a uni-axial BEC with $\lambda_{a}=\lambda_{b}$, the integral
$J_{ab}$ can be solved analytically, it reads
\begin{eqnarray*}
J_{ab} & = & \frac{-\sqrt{\left(\lambda_{c}^{2}+W\right)}\left(5\lambda_{a}^{2}-2\lambda_{c}^{2}+3W\right)}{4\left(\lambda_{a}^{2}+W\right)^{2}\left(\lambda_{a}^{2}-\lambda_{c}^{2}\right)^{2}}\\
 &  & +\frac{3}{8}\frac{\pi-2\arcsin\sqrt{\frac{\lambda_{c}^{2}+W}{\lambda_{a}^{2}+W}}}{\left(\lambda_{a}^{2}-\lambda_{c}^{2}\right)^{5/2}}.
\end{eqnarray*}
The sum rule for the double index integrals reads
\[
\begin{split}\frac{2}{\sqrt{\left(\lambda_{x}^{2}+W\right)\left(\lambda_{y}^{2}+W\right)\left(\lambda_{z}^{2}+W\right)}} & \frac{1}{\left(\lambda_{a}^{2}+W\right)}\\
\\
=2J_{aa}+J_{ax}+J_{ay}+J_{az}, & ,\quad a\in\left\{ x,y,z\right\} ,
\end{split}
\]
so that we have for example
\[
J_{zz}=\frac{2-J_{xz}-J_{yz}}{3\sqrt{\left(\lambda_{x}^{2}+W\right)\left(\lambda_{y}^{2}+W\right)\left(\lambda_{z}^{2}+W\right)}\left(\lambda_{z}^{2}+W\right)}.
\]
In the spherical case with $\lambda_{x}=\lambda_{y}=\lambda_{z}$
the result of the integral reads
\[
J_{xx}=J_{yy}=J_{zz}=\frac{2}{5\left(\lambda_{x}^{2}+W\right)^{5/2}}.
\]
With that we have everything at hand to calculate the mirror potential.
The index integral and their algebraic properties are also discussed in
\cite{Sapina} and \cite{Chandrasekhar}.

\end{document}